\documentclass[3p,times]{elsarticle}

\usepackage{ecrc}
\usepackage{amsmath}
\usepackage{color,hyperref} 
\usepackage{amssymb,dsfont}

\volume{00}

\firstpage{1}

\journalname{Physica A}

\runauth{}

\jid{procs}

\jnltitlelogo{Physica A}

\usepackage{amssymb}

\usepackage[figuresright]{rotating}

\begin{document}

\begin{frontmatter}

\dochead{}

\title{Area under subdiffusive random walks}

\author{V. M\'endez, R. Flaquer-Galm\'es and J. Cristín}

\address{Grup de F\'{\i}sica Estad\'{\i}stica, Departament de F\'{\i}sica. Facultat de Ci\`{e}ncies, Universitat Aut\`{o}noma de Barcelona, 08193 Barcelona, Spain}

\begin{abstract}
We study the statistical properties of the area and the absolute area under the trajectories of subdiffusive random walks. Using different frameworks to describe subdiffusion (as the scaled Brownian motion, fractional Brownian motion, the continuous-time random walk or the Brownian motion in heterogeneous media), we compute the first two moments, the ergodicity breaking parameter for the absolute area and infer a general scaling for the probability density functions of these functionals. We discuss the differences between the statistical properties of the area and the absolute area for the different subdiffusion models and illustrate the experimental interest of our results. Our theoretical findings are supported by Monte Carlo simulations showing an excellent agreement.
\end{abstract}

%\begin{keyword}
%\end{keyword}

\end{frontmatter}

\section{Introduction}\label{sec:intro}

Anomalous transport with sublinear mean-squared displacement (MSD $\sim t^{\gamma}$, $0<\gamma<1$) is a hallmark of complex media, from crowded intracellular environments to porous and fractal substrates~\cite{Woringer2020Frontiers,Holmes2019BJ,Tan2018PRL,Fomin2011FDC,Zhang2024JMEP,Xu2025Fractals}. In living cells, macromolecular crowding, transient binding, and viscoelastic response yield subdiffusive motion of proteins, chromatin loci, and vesicles; at interfaces and boundaries these dynamics can produce nontrivial spatial organization (e.g., depletion layers) with functional consequences for signaling and reaction kinetics~\cite{Woringer2020Frontiers,Holmes2019BJ}. In porous or fractal media, tortuosity and multiscale geometry likewise lead to subdiffusive transport and modified effective parameters that are crucial to interpretation and upscaling~\cite{Fomin2011FDC,Zhang2024JMEP,Xu2025Fractals}.

A variety of stochastic models have been proposed to describe these observations. On the one hand, one can consider the mesoscopic description of continuous-time random walks (CTRWs) with heavy-tailed waiting times, which exhibits aging and weak ergodicity breaking \cite{MeKl01}; their scaling limits connect to fractional diffusion equations and admit both subordination and Langevin formulations in physical time \cite{CairoliBaule2015PRE}. On the other hand, one can also consider the microscopic description based on the Langevin equation as the  scaled Brownian motion (SBM) and the fractional Brownian motion (fBM). Both are  Gaussian yet nonergodic and nonstationary processes widely employed to model a huge variety of physical and biological phenomena.  For example, SBM
was used to describe fluorescence recovery after photobleaching in various settings \cite{Sa01}
as well as anomalous diffusion in various biophysical contexts \cite{Sz06,Wu08,Me14}. In other branches of
physics SBM was used to study the scaled voter model \cite{Ka23} and to model turbulent flows observed by Richardson \cite{Ri26}. Moreover, the diffusion of particles in granular gases with
relative speed dependent restitution coefficients follow SBM \cite{Bo15}. FBM in the subdiffusive regime ($0<H<1/2$) has been observed in various tracers in complex environments both in vivo and in vitro \cite{Ma09,Je11,Je12,Je13,Ta13}, but also for completely different stochastic processes such
as electronic network traffic \cite{Mi02} or financial time series \cite{Co98,Bi08}. In the superdiffusive
regime ($1/2<H<1$), positive increment correlations and single trajectory powerspectra
consistent with fBM were observed for the actively driven motion of endogenous
granules inside amoeba cells as well as for the motion of the amoeba themselves \cite{Re15,Kr19}. Another relevant  microscopic description of subdiffusion based on the Langevin equation is the Brownian motion in heterogeneous media (HBM) with space-dependent diffusion coefficients. It has been used successfully in several contexts, such as Richardson diffusion in turbulence \cite{Ri26}, mesoscopic transport in disordered porous media \cite{De12}, and diffusion on random fractal structures \cite{Lo09}, random walks in an inhomogeneous medium \cite{LaBa01,PeJi16,ReSh16} ,
chemical reactions \cite{Ga04}, diffusion (in momentum space) in laser cooling processes \cite{BaBo01}, dissipative particle dynamics \cite{FaGr16}, vortex-antivortex annihilations \cite{Br00}, biophysics \cite{PiHe16,BeMa17,Ku11} e.g. measurements of proteins' diffusivity in mammalian cells \cite{Ku11}, and modeling of $1/f$ noise \cite{KaAl09}.

To characterize these movement models beyond kinematic quantities (MSD and propagators), stochastic functionals of random walks sharpen inference by linking trajectories to time-averaged observables, first-passage, and occupation statistics for example, and have been intensively studied, especially when the random walk is a Brownian motion \cite{Ma05}. 
Recently, the study of stochastic functionals of anomalous diffusion gained interest \cite{Ba06,BaFlMe23,CaBa10,LeBa19}. Among the variety of stochastic functionals of physical significance, the area and the absolute area under the trajectory of a random walk have been widely studied, particularly for the standard Brownian motion \cite{Sh82,PeWe96,Ta93,MaBr02}. The area has also been studied when the random walk is a Brownian motion under a potential \cite{Sm24} or under resetting \cite{Ho19}. The area can be related to the random acceleration process \cite{Ch43,BaFi06}: if the acceleration of a particle is a white Gaussian process, then the particle's position is equivalent to the area of a Brownian motion in the overdamped limit. 

In this context, experimental techniques based on nuclear magnetic resonance (NMR) have often been used to study atomic or molecular motion, and the application of inhomogeneous magnetic fields has allowed the characterization of the trajectories of spin-bearing particles \cite{Gr07}. The macroscopic measured signal $E$ in NMR experiments can be written as $E(t)=\left\langle e^{i\varphi}\right\rangle$ where $\varphi=g\int _0^t B[x(\tau)]d\tau$ is the phase accumulated by each spin, $g$ is the gyromagnetic ratio, $B(x)$ is a spatially-inhomogeneous external magnetic field, and $x(\tau)$ is the trajectory of each particle. So that $E(t)$ encodes information regarding the motion of the particles. 
Common choices of the magnetic field are $B\sim x$ or $B\sim x^2$ (see Eqs. 62 and 64 in Ref. \cite{Gr07}). For the first case, the measured signal $E(t)$ corresponds to the area under the trajectory. As mentioned above, these trajectories are often subdiffusive in nature, and the theoretical prediction for the measured signal may be useful to characterize the motion of the particles.

In this work, we study the statistical properties of two functionals of subdiffusive random walks: the area and the absolute area. To do this, we consider the different approaches to subdiffusion mentioned above: the subdiffusive-CTRW with heavy-tailed waiting times, SBM, fBM, and HBM with power-law spatially dependent diffusion coefficient. For the area, we compute the probability density function (PDF) and the first two moments while for the absolute area, we compute the first two moments, the ergodicity breaking parameter, and the scaling form of the PDF. Our analytical results have been tested with numerical simulations.

The paper is structured as follows: In Sec. \ref{sec:funct} we introduce the concept of stochastic functional, how to find the first two moments from the one-time and two-time PDFs, and how to compute the ergodicity breaking parameter from these moments. In Sec. \ref{sec:area}, we study the area under subdiffusive trajectories. We begin by finding general expressions for the first two moments in terms of the autocorrelation function. Then, we specify these results to subdiffusive-CTRW, SBM, fBM, and HBM. In Sec. \ref{sec:absolute_area}, we do the same for the absolute area, and we further compute the ergodicity breaking parameter, and compare the results obtained form the different models. We include some appendices to detail calculations. In Sec. \ref{sec:PDFs} we find the scaling form of the PDFs of the area and the absolute area and, moreover, for the area under Gaussian process we find the exact PDF. Our theoretical results are checked with numerical simulations and compared with previous results. Finally, the conclusions are presented in Sec. \ref{sec:conclusions}.

\section{Stochastic functionals and ergodic properties}\label{sec:funct}
Consider the stochastic functional
\begin{equation}
    Z(t)=\int_{0}^{t}U[x(\tau)]d\tau
\end{equation}
where $U[x(\tau)]$ is function of the stochastic trajectory $\{x(\tau); 0\leq \tau\leq t\}$ of a particle  which is initially at $x_0$, this is, $x(t=0)=x_0$. The mean value of the process $Z(t)$ if the motion started at $x=x_0$ is
\begin{eqnarray}
    \left\langle Z(t|x_0)\right\rangle &=&\int_{0}^{t}\left\langle U[x(\tau)]\right\rangle d\tau\nonumber\\
    &=&\int_{0}^{t}d\tau\int_{-\infty}^{\infty}dxU(x)P(x,\tau|x_0),
    \label{Z}
\end{eqnarray}
where $P(x,t|x_0)$ is the probability of finding the particle at position $x$ at time $t$ if it was initially at $x_0$. Analogously, the mean square values of the process $Z(t)$ is

\begin{eqnarray}
    \left\langle Z(t|x_0)^{2}\right\rangle &=&\left\langle \left(\int_{0}^{t}U[x(\tau)]d\tau\right)^{2}\right\rangle =2!\int_{0}^{t}d\tau_{2}\int_{0}^{\tau_{2}}d\tau_{1}\left\langle U[x(\tau_{1})]U[x(\tau_{2})]\right\rangle \nonumber\\
    &=&2!\int_{0}^{t}d\tau_{2}\int_{0}^{\tau_{2}}d\tau_{1}\int_{-\infty}^{\infty}dx_{2}\int_{-\infty}^{\infty}dx_{1}U(x_{1})U(x_{2})P(x_{2},\tau_{2};x_{1},\tau_{1}|x_0).
    \label{z21}
\end{eqnarray}
From these two mean values we can derive an expression for the ergodicity breaking parameter which will be useful to study the ergodic properties of $Z(t)$. 
Consider an observable $\mathcal{O}[x(\tau)]$, a function of the trajectory $\{x(\tau); 0\leq \tau\leq t\}$. Since $x(\tau)$ is stochastic in nature, the observable $\mathcal{O}[x(\tau)]$ will also be fluctuating between the realizations. An observable of the random walk is said to be ergodic if the ensemble average equals the time average $\left\langle \mathcal{O}\right\rangle =\ensuremath{\overline{\mathcal{O}}}$ in the long time limit. This means that if $\mathcal{O}[x(\tau)]$ is ergodic, then its time average $\ensuremath{\overline{\mathcal{O}}}$ is not a random variable. As a consequence, the limiting PDF of $\ensuremath{\overline{\mathcal{O}}}$, namely $P(\overline{\mathcal{O}},t)$, is a Dirac delta function:
\begin{eqnarray}
P(\overline{\mathcal{O}},t\to\infty)=\delta\left(\overline{\mathcal{O}}-\left\langle \mathcal{\overline{O}}\right\rangle \right).
     \label{lpdf}
 \end{eqnarray} 
 If the observable is integrable with respect to the density $P(x,t|x_0)$, then the ensemble average is given by
\begin{equation}
    \left\langle \mathcal{O}[x(t)]\right\rangle =\int_{-\infty}^{\infty}\mathcal{O}[x]P(x,t|x_0)dx.
\end{equation}
The time average of $\mathcal{O}[x(t)]$ is defined as
\begin{equation}
    \ensuremath{\overline{\mathcal{O}[x(t)]}=}\frac{1}{t}\int_{0}^{t}\mathcal{O}[x(\tau)]d\tau.
    \label{tav}
\end{equation}
For non-ergodic observables, since $\overline{\mathcal{O}}$ is random, its variance $\textrm{Var}(\overline{\mathcal{O}})$ is non-zero in the long time limit. Otherwise, for an ergodic observable $\textrm{Var}(\overline{\mathcal{O}})=0$ in the long time limit. Keeping this in mind, the ergodicity breaking parameter EB is defined as
\begin{eqnarray}
\textrm{EB}=\lim_{t\to\infty}\frac{\textrm{Var}(\overline{\mathcal{O}})}{\left\langle \overline{\mathcal{O}}\right\rangle ^{2}}=\lim_{t\to\infty}\frac{\left\langle \overline{\mathcal{O}}^{2}\right\rangle -\left\langle \overline{\mathcal{O}}\right\rangle ^{2}}{\left\langle \overline{\mathcal{O}}\right\rangle ^{2}}.
  \label{EB}
\end{eqnarray}
For ergodic observables, one should have $\textrm{EB}= 0$. 

In the examples below we consider the observable $\mathcal{O}[x(t)]=U[x(t)]$ so that the time average of the observable is from \eqref{tav}
\begin{eqnarray}
    \ensuremath{\overline{\mathcal{O}[x(t)]}=}\frac{1}{t}\int_{0}^{t}U[x(\tau)]d\tau=\frac{Z(t)}{t},
\end{eqnarray}
and so
\begin{eqnarray}
 \ensuremath{\left\langle \overline{\mathcal{O}}\right\rangle =}\frac{\left\langle Z(t)\right\rangle }{t},\quad\ensuremath{\left\langle \overline{\mathcal{O}}^{2}\right\rangle =}\frac{\left\langle Z(t)^{2}\right\rangle }{t^{2}}.   
 \label{o}
\end{eqnarray}
Finally, from \eqref{EB} we can find the EB in terms of the first two moments of the functional
\begin{eqnarray}
    \textrm{EB}=\frac{\left\langle Z(t)^{2}\right\rangle }{\left\langle Z(t)\right\rangle ^{2}}-1
    \label{EB2}
\end{eqnarray}
as $t\to \infty$. In this work we are interested in two specific functionals, the area and the absolute area under a random path. In Fig. \label{fig1} we illustrate the concepts of both areas.

\begin{figure}[h!]
    \centering
    \includegraphics[width=0.8\linewidth]{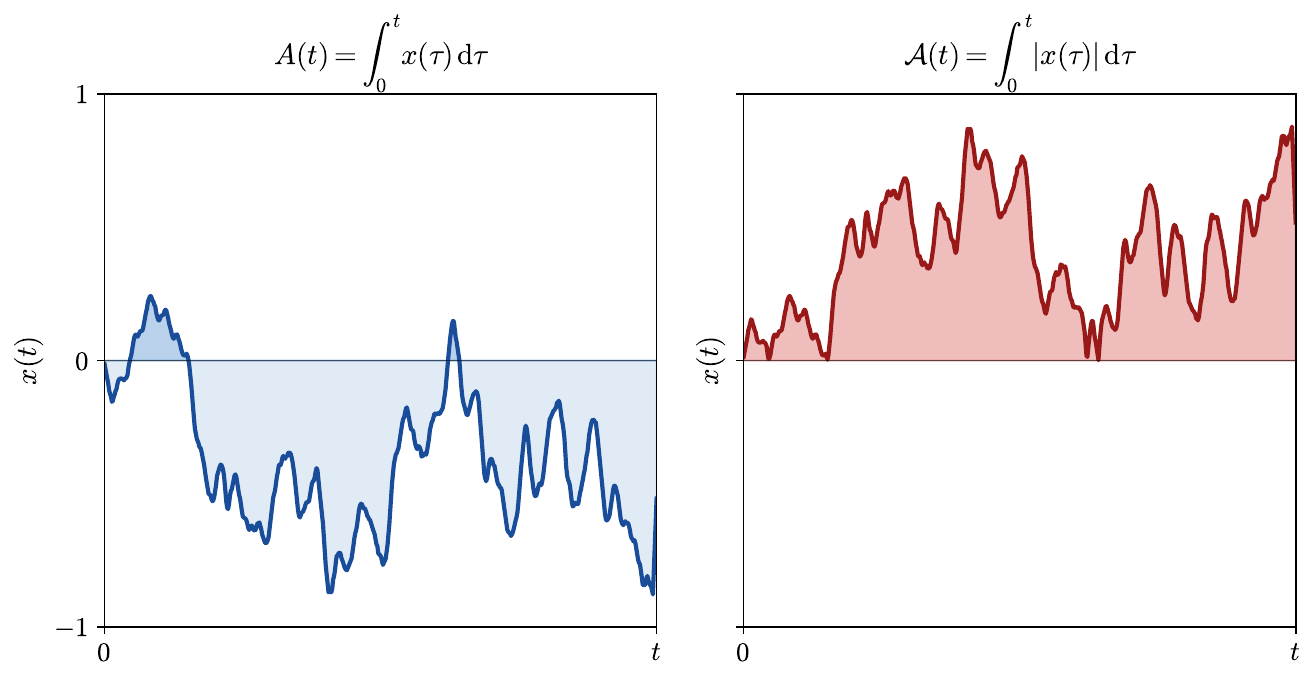}
    \caption{Schematic picture of the area and absolute area under a random path up to time $t$. }
    \label{fig1}
\end{figure}

\section{Area}\label{sec:area}
In this section we find the first two moments of the area swept by a subdiffusive random walker, while in Sec. \ref{sec:PDFs} we will analyze its PDFs. By definition, the area under the trajectory $\{x(\tau); x(0)=0, 0\leq \tau\leq t\}$ of the particle is a stochastic process defined by
\begin{eqnarray}
    A(t)=\int _0^t x(\tau)d\tau.
\end{eqnarray}
It is then a stochastic functional of the process $x(\tau)$ with $U(x)=x$. We consider that the random walk is isotropic, so that $\left\langle x(\tau)\right\rangle =0$ and in consequence the first moment is
\begin{eqnarray}
    \left\langle A(t)\right\rangle =0.
\end{eqnarray}
The second moment of the area can be written in general, from Eq. \eqref{z21}, as
\begin{eqnarray}
   \left\langle A(t)^{2}\right\rangle =\int_{0}^{t}dt_{1}\int_{0}^{t}dt_{2}\left\langle x(t_{1})x(t_{2})\right\rangle=2\int_{0}^{t}dt_{1}\int_{0}^{t_1}dt_{2}C(t_{1},t_{2})
    \label{eqcA}
\end{eqnarray}
where $C(t_{1},t_{2})\equiv\left\langle x(t_{1})x(t_{2})\right\rangle$. The second moment can be found from the autocorrelation function of the particle's position $C(t_1,t_2)$. Since in this case the mean value of the area is zero, it makes no sense to compute here the ergodicity breaking  for this observable. In what follows, we present the expression for the second moment of the area \eqref{eqcA} for the specific examples of subdiffusion studied in this work.

\subsection{Area under the trajectory of a Brownian motion}
For the standard Brownian motion the autocorrelation of the particle's position is given by $C(t_2,t_1)=2D\min (t_2,t_1)$. Since in the integrals of Eq. \eqref{aa2} we have $0<t_1<t_2<t$, then 
$C(t_2,t_1)=2Dt_1=C(t_1,t_1)$ and $C(t_2,t_2)=2Dt_2$. Finally from \eqref{eqcA} 
\begin{eqnarray}
\left\langle A(t)^{2}\right\rangle &=&2D\int_{0}^{t}dt_{1}\left[\int_{0}^{t_{1}}\min(t_{1},t_{2})dt_{2}+\int_{t_{1}}^{t}\min(t_{1},t_{2})dt_{2}\right]\nonumber\\
&=&2D\int_{0}^{t}dt_{1}\left[\frac{t_{1}^{2}}{2}+t_{1}(t-t_{1})\right]=\frac{2Dt^{3}}{3}.
   \label{a2bm}
\end{eqnarray}
This result was originally derived by L\'evy \cite{Le40b} and has been found  as the solution of the Obukhov model for turbulent flows \cite{Ob59,BaFi06}.

\subsection{Area under the trajectory of a Brownian motion with time-dependent diffusivity}
Consider now that the trajectory $\{x(\tau); x(0)=0, 0\leq \tau\leq t\}$ obeys the Langevin equation with time-dependent diffusivity 
\begin{equation}
    \frac{dx(t)}{dt}=\sqrt{2D(t)}\xi(t)
    \label{LE}
\end{equation}
where $\xi (t)$ is white Gaussian noise with zero mean and unit amplitude $\left\langle \xi(t)\xi(t')\right\rangle =\delta(t-t')$. Different temporal dependencies of the diffusion coefficient have been considered in the physical literature \cite{Yu16,Co25,He25}.  Since $\xi(t)$ is Gaussian, so is $x(t)$ and its autocorrelation reads
\begin{eqnarray}
    C(t_{1},t_{2})=2\int_{0}^{t_{1}}\sqrt{D(\tau_{1})}d\tau_{1}\int_{0}^{t_{2}}\sqrt{D(\tau_{2})}\left\langle \xi(\tau_{1})\xi(\tau_{2})\right\rangle d\tau_{2}=2\int_{0}^{\min(t_{1},t_{2})}D(\tau)d\tau,
    \label{C12}
\end{eqnarray}
i.e, the MSD reads 
\begin{eqnarray}
    \left\langle x(t)^{2}\right\rangle =2\int_{0}^{t}D(\tau)d\tau.
    \label{msd1}
\end{eqnarray}
From \eqref{eqcA} and \eqref{C12} we find, after some calculations,
\begin{eqnarray}
    \left\langle A(t)^{2}\right\rangle =4\int_{0}^{t}dt'(t-t')\int_{0}^{t'}D(t'')dt''=2\int_{0}^{t}(t-t')^{2}D(t')dt',
    \label{A2dt}
\end{eqnarray}
where we have interchanged the order of integration in the last equality. Eq. \eqref{A2dt} is interesting because it allows to compute the variance of the area under the trajectory of a Brownian particle with time-dependent diffusivity just in terms of the diffusivity. 
A particularly relevant example of time-dependent diffusivity consists in assuming the dependence 
\begin{eqnarray}
    D(t) = \alpha K t^{\alpha-1}
    \label{Dsbm}
\end{eqnarray}
with $\alpha \in (0,2)$, which corresponds to the Scaled Brownian Motion (SBM), for which the MSD is given by
\begin{eqnarray}
    \left\langle x(t)^{2}\right\rangle =2Kt^\alpha.
    \label{msdsbm}
\end{eqnarray}
In this case, the  autocorrelation function of $x(t)$ reads
\begin{eqnarray}
C(t_{1},t_{2})=2K\left[\min(t_{1},t_{2})\right]^{\alpha},
    \label{cmsd}
\end{eqnarray}
which can be derived integrating \eqref{LE} with \eqref{Dsbm}. Combining Eqs. \eqref{C12} and \eqref{Dsbm}, we find 
\begin{eqnarray}
    \left\langle A(t)^{2}\right\rangle =2\alpha K\int_0^t (t-t')^2 (t')^{\alpha-1} dt'=\frac{4Kt^{\alpha+2}}{(\alpha+1)(\alpha+2)}.
    \label{A2sbm}
\end{eqnarray}
Note that for $\alpha=1$ Eq. \eqref{A2sbm} reduces to \eqref{a2bm} as expected.

%\begin{figure}[h]
%    \centering
%    \includegraphics[width=0.99\linewidth]{Fig1_A2_SBM_fBM.png}
%    \caption{Second moment of the area $\langle A^{2}(t) \rangle$ as a function of time $t$. (a) Scaled Brownian motion (SBM). The lines correspond to the theoretical prediction given by Eq.~\eqref{A2sbm}. (b) Fractional Brownian motion (FBM). The lines correspond to the theoretical prediction given by Eq.~\eqref{A2fbm}.}%, as a visual guide. b) Probability distribution of $P(A)$ vs $A$ at a given time $t$. The dotted lines correspond to equation (xx) in the main text. c) a) Normalized $\left\langle \mathcal{A}^2(t) \right\rangle / \left\langle \mathcal{A}^2_{Th}(t) \right\rangle^2$ vs time $t$, to observe the convergence towards the theoretical prediction. The dotted line corresponds to $1$, as a visual guide. d) Ergodicity Breaking (EB) parameter vs time $t$. The dotted lines correspond to equation (xx) in the main text.}
%    \label{fig:A2_g}
%\end{figure}

\begin{figure}[h]
    \centering
    \includegraphics[width=0.99\linewidth]{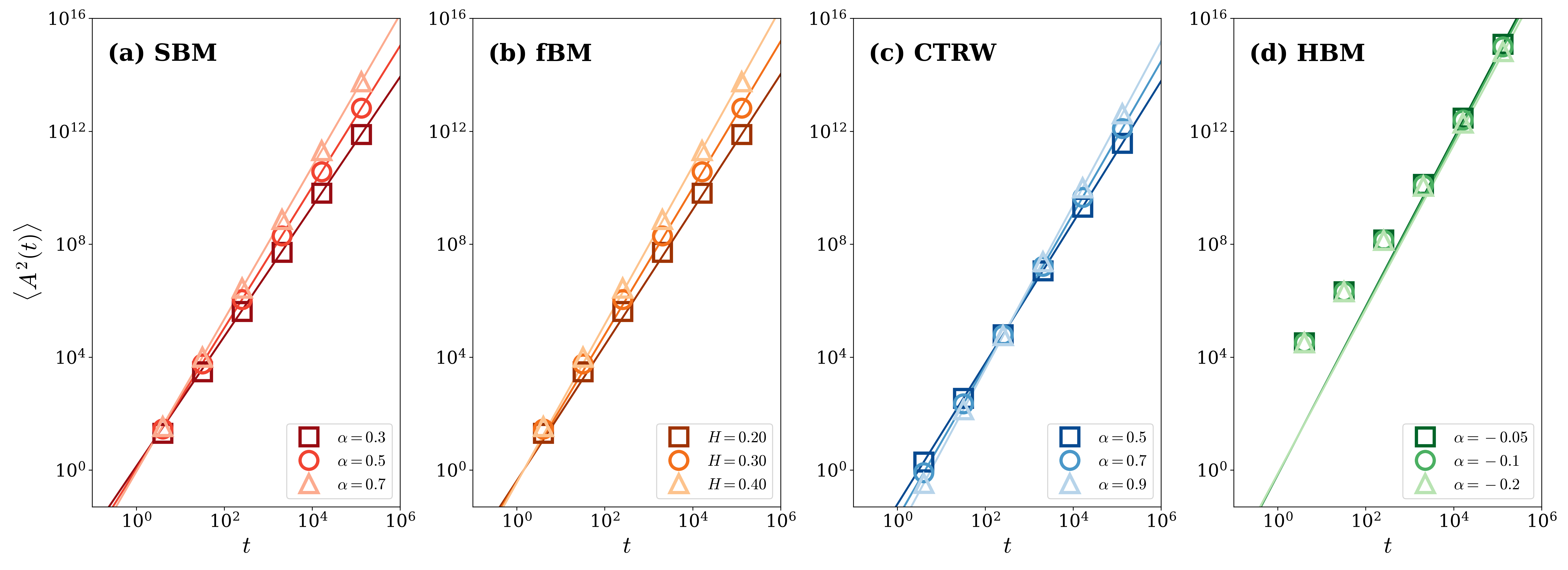}
    \caption{Second moment of the area $\langle A^{2}(t) \rangle$ as a function of time $t$. (a) Scaled Brownian motion (SBM). The lines correspond to the theoretical prediction given by Eq.~\eqref{A2sbm}. (b) Fractional Brownian motion (fBM). The lines correspond to the theoretical prediction given by Eq.~\eqref{A2fbm}. (c) Continuous Time Random Walk (CTRW). The lines correspond to the theoretical prediction given by Eq.~\eqref{a2sct}. (d) Heterogeneous Brownian motion (HBM). The lines correspond to the theoretical prediction given by Eq.~\eqref{A2dx}. The parameters used in the simulations are listed in \ref{app:simulations}.}
    \label{fig:A2}
\end{figure}

In Fig. \ref{fig:A2}(a) we verify the analytical results for $\langle A^{2}(t) \rangle$ with numerical simulations (details of the simulation algorithm can be found in \ref{app:simulations}).

\subsection{Area under the trajectory of a fractional Brownian motion}
Next we consider that the trajectory $\{x(\tau); x(0)=0, 0\leq \tau\leq t\}$ follows a fractional Brownian motion (fBM). It is a self-similar Gaussian process with zero mean and autocorrelation
\begin{eqnarray}
C(t_1,t_2)=D_{H}\left(t_{1}^{2H}+t_{2}^{2H}-|t_{1}-t_{2}|^{2H}\right)
\label{C}
\end{eqnarray}
with $H\in (0,1)$, so that $\left\langle x(t)^{2}\right\rangle=2D_Ht^{2H}$. 
The fBM was first introduced by Kolmogorov \cite{Ko40} and later studied by Mandelbrot and van Ness \cite{Ma68}. 
From the definition \eqref{C} and Eq. \eqref{eqcA}, the variance of the area is
\begin{eqnarray}
\left\langle A(t)^{2}\right\rangle &=&\int_{0}^{t}dt_{1}\int_{0}^{t}dt_{2}D_{H}\left(t_{1}^{2H}+t_{2}^{2H}-|t_{1}-t_{2}|^{2H}\right)\nonumber\\
&=&\frac{2D_{H}}{2H+1}t^{2H+2}-2D_{H}\int_{0}^{t}dt_{1}\int_{0}^{t_{1}}dt_{2}(t_{1}-t_{2})^{2H}
=\frac{D_{H}t^{2H+2}}{H+1}.
  \label{A2fbm}
\end{eqnarray}
In Fig. \ref{fig:A2}(b) we verify the analytical results for $\langle A^{2}(t) \rangle$ with numerical simulations (details of the simulation algorithm can be found in \ref{app:simulations}).

\subsection{Area under the trajectory of a subdiffusive-CTRW}
Consider a CTRW in which the particle performs instantaneous jumps. Each jump is followed by a rest of duration $\tau$ drawn from the waiting time PDF $\varphi (\tau)$. If we consider the waiting time PDF
\begin{eqnarray}
   \varphi(t)=\frac{t^{\alpha-1}}{\tau^{\alpha}}E_{\alpha,\alpha}\left(-\frac{t^{\alpha}}{\tau^{\alpha}}\right), 
   \label{ml}
\end{eqnarray} 
where $E_{\alpha,\beta}(\cdot)$ is the Mittag-Leffler function, in the continuum limit the motion of the particle is subdiffusive \cite{MeKl01}. 
The second moment of the area under a trajectory of a subdiffusive-CTRW is (see \ref{app:a2sct} for detailed calculations)
\begin{eqnarray}
    \left\langle A(t)^2\right\rangle =\frac{4K_{\alpha}}{\Gamma\left(3+\alpha\right)}t^{2+\alpha}
    \label{a2sct}
\end{eqnarray}
where $K_\alpha =D/\tau ^{\alpha-1}$. This result has been also obtained in Ref. \cite{CaBa10} by projecting the fractional Feynman-Kac equation and solving the resulting equation for the second moment. In Fig. \ref{fig:A2}(c) we verify the analytical results for $\langle A^{2}(t) \rangle$ with numerical simulations (details of the simulation algorithm can be found in \ref{app:simulations}).

\subsection{Area under the trajectory of a HBM}

Let $x(t)$ be the position of a Brownian particle moving in a one-dimensional heterogeneous media. It is a stochastic process evolving according the overdamped Langevin equation with a space-dependent diffusion coefficient, namely, 
\begin{eqnarray}
        \frac{dx(t)}{dt}=\sqrt{2D(x)}\xi(t)
    \label{le2}
\end{eqnarray}
where $\xi(t)$ is a Gaussian noise with zero mean and autocorrelation $\left\langle \xi(t)\xi(t')\right\rangle =\delta(t-t')$. 
We assume the power law dependence $D(x)=D_0|x|^\alpha$ with $\alpha<2$  \cite{Sr07,Che13,LeBa19,Radice23,Si22,WaDeCh19}. Hence, the Langevin equation for a Brownian particle with a diffusion coefficient with power-law dependence on space is
\begin{eqnarray}
\frac{dx(t)}{dt}=\sqrt{2D_0}|x|^{\alpha/2}\xi(t)
    \label{led2}.
\end{eqnarray}
To integrate \eqref{led2} we need to consider the appropriate interpretation of the stochastic calculus. The simplest option consists in considering the Stratonovich interpretation. In this interpretation it is possible to perform a change of variables which transforms the Langevin Eq.  \eqref{led2} into the Langevin equation for Brownian motion. Introducing the new variable
\begin{eqnarray}
    y(x)=\int\frac{dx}{\sqrt{2D(x)}}=\frac{\sqrt{2/D_{0}}}{2-\alpha}|x|^{\frac{2-\alpha}{2}}\textrm{sign}(x),
    \label{y(x)}
\end{eqnarray}
the Langevin Eq. \eqref{le2} becomes $dy(t)/dt=\xi(t)$ so that $y(t)$ is a Wiener process with the known Gaussian PDF for the initial condition
$y(0) = 0$, namely,
\begin{eqnarray}
    P(y,t)=\frac{1}{\sqrt{2\pi t}}\exp\left(-\frac{y^{2}}{2t}\right).
    \label{oty}
\end{eqnarray}
Returning to the $x$-variable, the one time PDF reads
\begin{eqnarray}
    P(x,t)=\frac{1}{2|x|^{\alpha/2}\sqrt{\pi D_{0}t}}\exp\left(-\frac{|x|^{2-\alpha}}{(2-\alpha)^{2}D_{0}t}\right)
    \label{pdfdx}
\end{eqnarray}
from which the MSD follows as
\begin{eqnarray}
    \left\langle x^{2}(t)\right\rangle =\int_{-\infty}^{\infty}x^{2}P(x,t)dx=\frac{(2-\alpha)^{\frac{4}{2-\alpha}}}{\sqrt{\pi}}\Gamma\left(\frac{6-\alpha}{2(2-\alpha)}\right)\left(D_{0}t\right)^{\frac{2}{2-\alpha}}
\end{eqnarray}
which had been reported in \cite{Che13}. Thus, for $\alpha<0$ we find subdiffusion, and superdiffusion for $\alpha>0$. Brownian motion emerges for $\alpha=0$, and $\alpha=1$ produces ballistic motion. The diffusion becomes increasingly fast when $\alpha$ tends to 2.
Again, the second moment of the area can be computed from Eq. \eqref{eqcA} where we refer to \ref{app:A2dx} for the details on the calculations. We finally obtain 
\begin{eqnarray}
    \left\langle A(t)^{2}\right\rangle =2D_{0}^{\frac{2}{2-\alpha}}\mathcal{B}(\alpha)t^{\frac{2(3-\alpha)}{2-\alpha}}
    \label{A2dx}
\end{eqnarray}
where
\begin{eqnarray*}
  \mathcal{B}(\alpha)
  =\frac{(2-\alpha)^{\frac{6-\alpha}{2-\alpha}}}{\pi}\Gamma\left(\frac{3-\alpha}{2-\alpha}\right)^{2}\frac{\Gamma\left(\frac{8-3\alpha}{4-2\alpha}\right)\Gamma\left(\frac{10-3\alpha}{4-2\alpha}\right)}{(3-\alpha)\Gamma\left(\frac{9-3\alpha}{2-\alpha}\right)}{}_{3}F_{2}\left(\frac{3-\alpha}{2-\alpha},\frac{3-\alpha}{2-\alpha},\frac{8-3\alpha}{4-2\alpha};\frac{3}{2},\frac{9-3\alpha}{2-\alpha};1\right).
\end{eqnarray*}
Notice that by setting $\alpha=0$ we recover \eqref{a2bm}. It is important to highlight that the previous result depends on the interpretation chosen for the stochastic calculus. In the Itô interpretation the results (even the exponent of time) would be different. In Fig. \ref{fig:A2}(d) we compare this result with numerical simulations showing a good agreement. The analytical result \eqref{A2dx} is exact but in Fig. \ref{fig:A2}(d) we observe that the agreement with simulations occurs from $t\sim10^4$ onward. This is due to the slow numeric convergence in the HBM case. Details of the simulation algorithm can be found in \ref{app:simulations}.

\section{Absolute Area}\label{sec:absolute_area}
In this section we find the first two moments of the absolute area swept by a subdiffusive random walker.
It is defined as the area below the absolute value of the trajectory, i.e., it is a stochastic functional with $U(x)=|x|$:
\begin{eqnarray}
    \mathcal{A}(t)=\int_0^t |x(\tau)|d\tau.
    \label{aad}
\end{eqnarray}
It was first studied in the context of economics \cite{Ci75} and later extensively in probability \cite{Sh82,Ta93}. The same functional was also studied in the context of electron-electron and phase coherence in one dimensional weakly disordered quantum wire \cite{Al82}. In this case our method to compute the first two moments from \eqref{Z} and \eqref{z21} is specially useful since the solution of the corresponding FK equation is extremely difficult, and one cannot get a closed form equation for the moments from it. To compute the mean value of $\mathcal{A}(t)$, we need the expression for $P(x,t)$. From \eqref{Z} and \eqref{aad}, since the underlying random walk is isotropic,
\begin{eqnarray}
    \left\langle \mathcal{A}(t)\right\rangle =\int_{0}^{t}\left\langle |x(\tau)|\right\rangle d\tau=\int_{0}^{t}d\tau\int_{-\infty}^{\infty}|x|P(x,\tau)dx=2\int_{0}^{t}d\tau\int_{0}^{\infty}xP(x,\tau)dx.
    \label{aa1m}
\end{eqnarray}
For the second moment, from \eqref{z21} and \eqref{aad} it follows that
\begin{eqnarray}
    \left\langle \mathcal{A}(t)^{2}\right\rangle =8\int_{0}^{t}d\tau_{2}\int_{0}^{\tau_{2}}d\tau_{1}\int_{0}^{\infty}x_{2}dx_{2}\int_{0}^{\infty}x_{1}dx_{1}P(x_{2},\tau_{2};x_{1},\tau_{1}).
    \label{aeqcA}
\end{eqnarray}
Below, we apply these general formulas to the specific examples of subdiffusion considered above for the area.

\subsection{Absolute area under Gaussian processes}
Consider that the particle's position follows a Gaussian process. The expression for $P(x,t)$  can be derived from the characteristic functional of $x(t)$, which for a Gaussian process is given by \cite{vK92}
\begin{eqnarray}
    \left\langle e^{i\int_{-\infty}^{\infty}f(\tau)x(\tau)d\tau}\right\rangle =\exp\left[-\frac{1}{2}\int_{-\infty}^{\infty}dt_{1}\int_{-\infty}^{\infty}dt_{2}f(t_{1})f(t_{2})C(t_{1},t_{2})\right]
    \label{cf}
\end{eqnarray}
where $f(t)$ is an arbitrary integrable function. Specifically, by taking $f(\tau)=k\delta(\tau-t)$ to get $P(k,t)=\left\langle e^{ikx(t)}\right\rangle =\exp[-k^2C(t,t)/2]$ after Fourier inversion yields
\begin{eqnarray}
   P(x,t)= \frac{e^{-\frac{x^{2}}{2C(t,t)}}}{\sqrt{2\pi C(t,t)}}.
\end{eqnarray}
Inserting this into \eqref{aa1m} and integrating over $x$ we find
\begin{eqnarray}
    \left\langle \mathcal{A}(t)\right\rangle =\sqrt{\frac{2}{\pi}}\int_{0}^{t}\sqrt{C(\tau,\tau)}d\tau= \sqrt{\frac{2}{\pi}}\int_{0}^{t}\sqrt{\langle x^2(\tau)\rangle}d\tau.
    \label{aa1}
\end{eqnarray}
This is an interesting result because it is an expression for the mean absolute area under a Gaussian process in terms of its MSD.  
To compute the mean value of $\mathcal{A}(t)^2$ we need the two-time PDF $P(x_{2},\tau_{2};x_{1},\tau_{1})$ for a Gaussian process. To get it we plug $f(\tau)=k_1\delta(\tau-t_1)+k_2\delta(\tau-t_2)$ into \eqref{cf} and obtain
\begin{eqnarray*}
    P(k_{2},t_{2};k_{1},t_{1})=\left\langle e^{ik_{1}x(t_{1})+ik_{2}x(t_{2})}\right\rangle &=&\exp\bigg[ 
    -\frac{k_{1}^{2}}{2}C(t_{1},t_{1})-\frac{k_{2}^{2}}{2}C(t_{2},t_{2})-k_{1}k_{2}C(t_{1},t_{2})\bigg]
\end{eqnarray*}
which after a double Fourier inversion yields
\begin{eqnarray}
P(x_{2},t_{2};x_{1},t_{1})=\frac{\exp\left[-\frac{C(t_{1},t_{1})x_{2}^{2}+C(t_{2},t_{2})x_{1}^{2}-2C(t_{1},t_{2})x_{1}x_{2}}{2\Phi(t_{1},t_{2})}\right]}{\pi\sqrt{2\Phi(t_{1},t_{2})}}
        \label{cftt}
\end{eqnarray}
where
$$
\Phi(t_{1},t_{2})=C(t_{1},t_{1})C(t_{2},t_{2})-C(t_{1},t_{2})^{2}.
$$
Now we are in position to compute the second moment from \eqref{aeqcA}. Using \eqref{cftt} and performing the integrals over $x_2$ and $x_1$ we find
\begin{eqnarray}
    \left\langle \mathcal{A}(t)^{2}\right\rangle =\frac{2}{\pi}\int_{0}^{t}dt_{2}\int_{0}^{t_{2}}dt_{1}C(t_{2},t_{1})\Psi(t_{2},t_{1})
    \label{aa2}
\end{eqnarray}
where
$$
\Psi(t_{2},t_{1})=\pi-2\arctan\left[\sqrt{\frac{C(t_{2},t_{2})C(t_{1},t_{1})}{C(t_{2},t_{1})^{2}}-1}\right]+2\sqrt{\frac{C(t_{2},t_{2})C(t_{1},t_{1})}{C(t_{2},t_{1})^{2}}-1}.
$$
This is also an interesting result because it is an expression for the mean square absolute area under a Gaussian process in terms of its autocorrelation function. Let us specify these general expressions for the same particular examples of Gaussian processes we studied for the area, for which the autocorrelation function of the particle's position can be found.

\subsubsection{Absolute area under a Brownian motion}
Regarding the absolute area of a Brownian motion, the mean values follows from \eqref{aa1} and \eqref{aa2} which yield
\begin{eqnarray}
    \left\langle \mathcal{A}(t)\right\rangle =\frac{4}{3}\sqrt{\frac{D}{\pi}}t^{\frac{3}{2}},\quad\left\langle \mathcal{A}(t)^{2}\right\rangle =\frac{3Dt^{3}}{4}.
    \label{aabm}
\end{eqnarray}
The characteristic function of $\mathcal{A}(t)$ and their moments were derived in \cite{Sh82,Ta93} but the exact Laplace inversion to find $P(\mathcal{A},t)_{\text{BM}}$ is not known. However, using the large deviation theory it has been possible to obtain some scaling properties of $P(\mathcal{A},t)_{\text{BM}}$ \cite{MaBr02}. The EB parameter can be readily found from and  \eqref{EB2} and \eqref{aabm}:
\begin{eqnarray}
    \textrm{EB}_{\textrm{BM}}=\pi\left(\frac{3}{4}\right)^{3}-1=0.325....
    \label{ebbm}
\end{eqnarray}
Since EB is always positive, the observable $|x(\tau)|$ is then non-ergodic. 

\subsubsection{Absolute area under SBM}
\begin{figure}[h]
    \centering
    \includegraphics[width=0.99\linewidth]{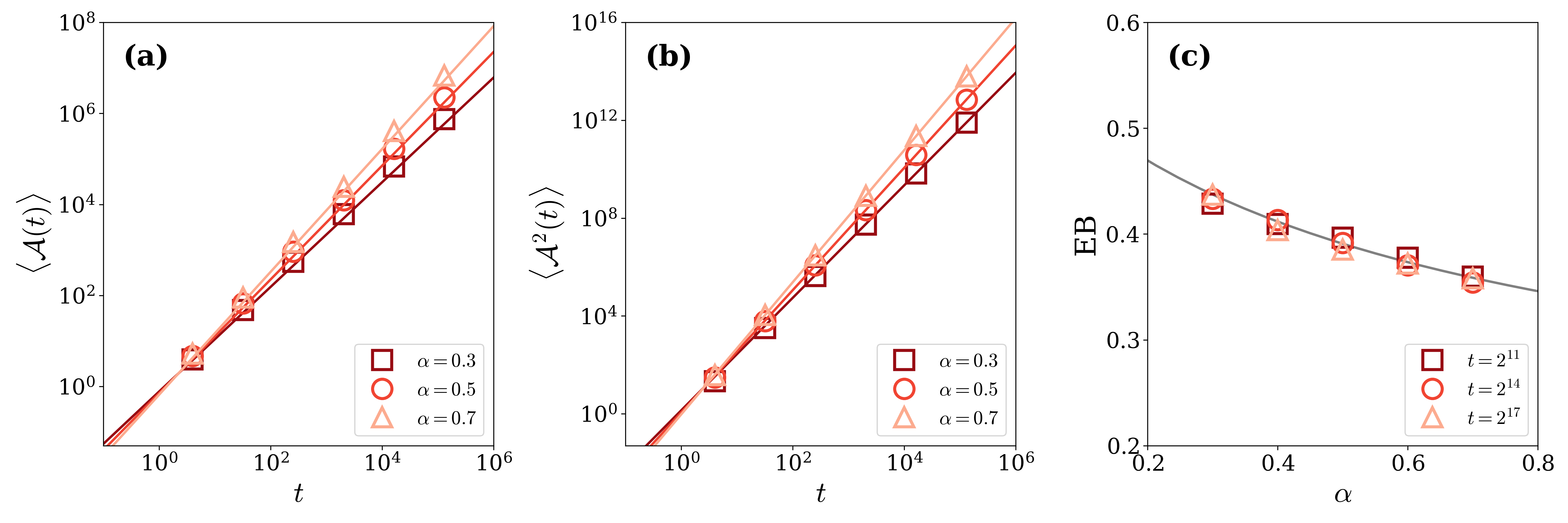}
    \caption{Scaled Brownian Motion (SBM). (a) First moment of the absolute area $\langle \mathcal{A}(t) \rangle$ as a function of time $t$. The lines correspond to the theoretical prediction given by Eq.~\eqref{aiaaaa}. (b) Second moment of the area $\langle \mathcal{A}^{2}(t) \rangle$ as a function of time $t$. The lines correspond to the theoretical prediction given by Eq.~\eqref{aiaaaa}. (c) Ergodicity breaking parameter EB as a function of the parameter $\alpha$.  The line corresponds to the theoretical prediction given by Eq.~\eqref{EBaasbm}. The simulations details are provided in \ref{app:simulations}.}
    \label{fig:SBM}
\end{figure}

To compute the first two moments of the absolute area we employ the formulas provided by Eqs. \eqref{aa1} and \eqref{aa2}. The first moment can be readily found from Eq. \eqref{cmsd} but the second moment needs some care. We detail the calculations in \ref{app:aiaaaa}. The final results are: 

\begin{eqnarray}
    \left\langle \mathcal{A}(t)\right\rangle =\frac{4}{2+\alpha}\sqrt{\frac{K}{\pi}}t^{1+\frac{\alpha}{2}},\quad\left\langle \mathcal{A}(t)^{2}\right\rangle =\frac{4Kt^{\alpha+2}}{(\alpha+1)(\alpha+2)}\left[1+\frac{1}{2\sqrt{\pi}}\frac{\Gamma\left(\frac{1}{2}+\frac{1}{\alpha}\right)}{\Gamma\left(2+\frac{1}{\alpha}\right)}\right].
    \label{aiaaaa}
\end{eqnarray}
The EB parameter follows easily from \eqref{EB2} as
\begin{eqnarray}
\textrm{EB}_{\textrm{SBM}}=\frac{\pi(\alpha+2)}{4(\alpha+1)}\left[1+\frac{1}{2\sqrt{\pi}}\frac{\Gamma\left(\frac{1}{2}+\frac{1}{\alpha}\right)}{\Gamma\left(2+\frac{1}{\alpha}\right)}\right]-1.
    \label{EBaasbm}
\end{eqnarray}
Note that setting $\alpha=1$, Eq. \eqref{EBaasbm} reduces to \eqref{ebbm}. EB decreases monotonically as $\alpha$ increases from 0 to 2. In Fig. \ref{fig:SBM} we show a comparison between the theoretical values. The agreement is excellent. In the numerical calculation of EB we can see how the convergence towards the asymptotic value in the long time limit is fast.

\subsubsection{Absolute area under fBM}
\begin{figure}[h]
    \centering
    \includegraphics[width=0.99\linewidth]{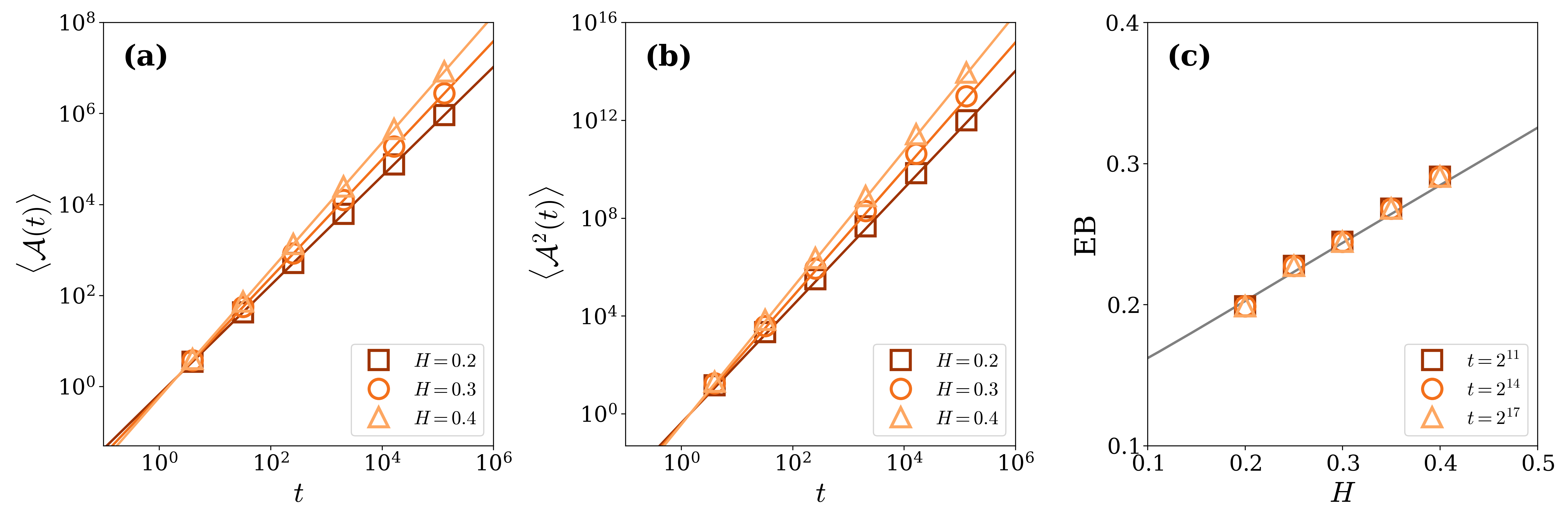}
    \caption{Fractional Brownian Motion (fBM). (a) First moment of the absolute area $\langle \mathcal{A}(t) \rangle$ as a function of time $t$. The lines correspond to the theoretical prediction given by Eq.~\eqref{a1a2aa}. (b) Second moment of the area $\langle \mathcal{A}^{2}(t) \rangle$ as a function of time $t$. The lines correspond to the theoretical prediction given by Eq.~\eqref{a1a2aa}. (c) Ergodicity breaking parameter EB as a function of the parameter $H$.  The line corresponds to the theoretical prediction given by Eq.~\eqref{EB_fbm}. The simulations details are provided in \ref{app:simulations}.}
    \label{fig:fBM}
\end{figure}

The mean value of the absolute area for a fBM follows easily from introducing the autocorrelation \eqref{C} into \eqref{aa1}. Inserting \eqref{C} into \eqref{aa2} and introducing the new variables $u=\tau_1/\tau_2$ and $v=\tau_2/t$ we write, after integrating over $v$, 
\begin{eqnarray}
    \left\langle \mathcal{A}(t)\right\rangle =\frac{2}{1+H}\sqrt{\frac{D_{H}}{\pi}}t^{1+H},\quad\left\langle \mathcal{A}(t)^{2}\right\rangle =\frac{D_{H}\lambda_{H}}{(H+1)\pi}t^{2+2H}
    \label{a1a2aa}
\end{eqnarray}
where the constant $\lambda_H$ is 
$$
\lambda_{H}=\pi-2\int_{0}^{1}\phi(u)\arctan\left[\frac{\Phi(u)}{\phi(u)}\right]du+2\int_{0}^{1}\Phi(u)du
$$
with
$$
\phi(u)=1+u^{2H}-(1-u)^{2H},\quad\Phi(u)=\sqrt{4u^{2H}-\phi(u)^{2}}.
$$
The ergodicity breaking parameter is obtained from \eqref{a1a2aa}
\begin{eqnarray}
    \textrm{EB}_{\textrm{fBM}}=\frac{1+H}{4}\lambda_{H}-1.
    \label{EB_fbm}
\end{eqnarray}
In Fig. \ref{fig:fBM} we compare the above results with numerical simulations showing an excellent agreement. The discussion of why EB behaves differently for SBM than for fBM is covered in Sec. \ref{sub:eb_parameter}.

\subsection{Absolute area under the trajectory of non-Gaussian process}
In this section we obtain the first two moment for the absolute area when the underlying random walk is not Gaussian. In consequence, we can not make use of Eqs. \eqref{aa1} and \eqref{aa2}; instead, we must use directly Eqs. \eqref{aa1m} and \eqref{aeqcA}.   
 
\subsubsection{Absolute area under a subdiffusive-CTRW}
\begin{figure}[h]
    \centering
    \includegraphics[width=0.99\linewidth]{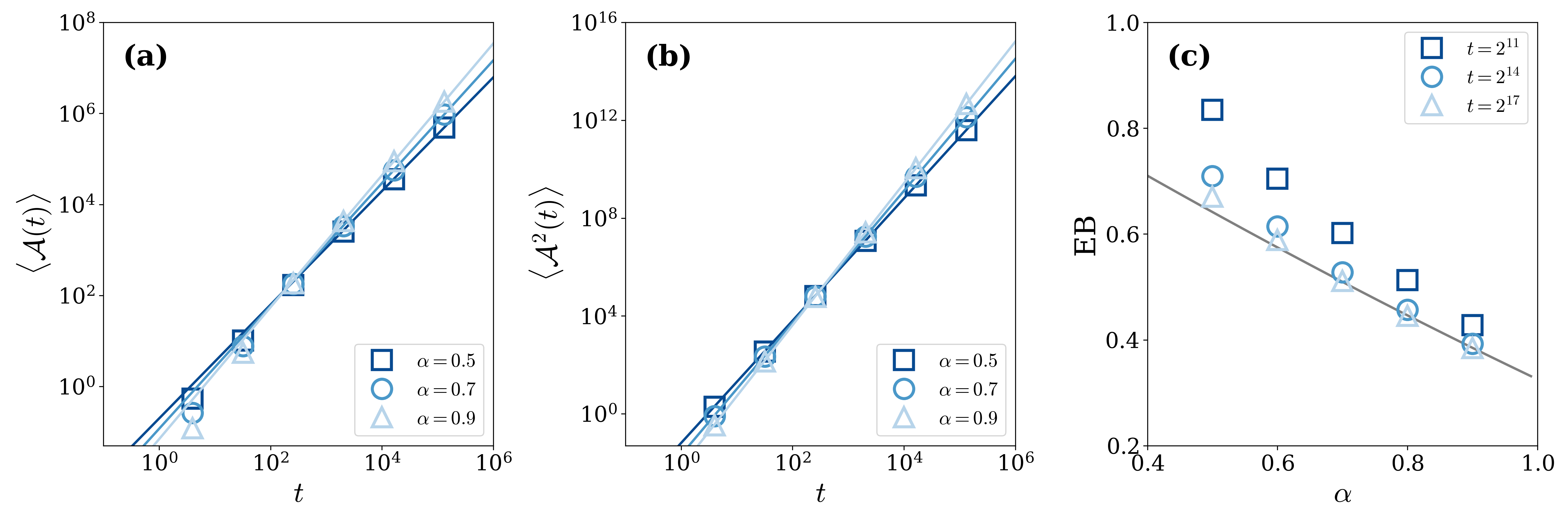}
    \caption{Continuous Time Random Walk (CTRW). (a) First moment of the absolute area $\langle \mathcal{A}(t) \rangle$ as a function of time $t$. The lines correspond to the theoretical prediction given by Eq.~\eqref{aa12}. (b) Second moment of the area $\langle \mathcal{A}^{2}(t) \rangle$ as a function of time $t$. The lines correspond to the theoretical prediction given by Eq.~\eqref{aa12}. (c) Ergodicity breaking parameter $EB$ as a function of the parameter $\alpha$.  The line corresponds to the theoretical prediction given by Eq.~\eqref{EB_ctrw}. The simulations details are provided in \ref{app:simulations}.}
    \label{fig:CTRW}
\end{figure}

Here we proceed as for the area, using the subordination between processes to find the one-time and two-time PDF of the subdiffusive-CTRW process. In \ref{app:aa12} we detail the calculations to obtain the first two moments of the absolute area. We obtain the following results: 
\begin{eqnarray}
    \left\langle \mathcal{A}(t)\right\rangle =\frac{\sqrt{K_{\alpha}}}{\Gamma\left(2+\frac{\alpha}{2}\right)}t^{1+\frac{\alpha}{2}},\quad \left\langle \mathcal{A}(t)^{2}\right\rangle =\frac{4K_{\alpha}}{\Gamma\left(3+\alpha\right)}\left(1+\frac{\alpha}{8}\right)t^{2+\alpha}
    \label{aa12}
\end{eqnarray}
and the EB parameter
\begin{eqnarray}
    \text{EB}_{\text{CTRW}}=\frac{(8+\alpha)\Gamma^{2}\left(2+\frac{\alpha}{2}\right)}{2\Gamma\left(3+\alpha\right)}-1.
        \label{EB_ctrw}
\end{eqnarray}
As for SBM the EB parameter decreases with $\alpha$. In Fig. \ref{fig:CTRW} we compare the above results with simulations. It is observed that the numerical convergence towards the actual value of EB is slower than for SBM or fBM. 

\subsubsection{Absolute area under HBM}
Again, we find the first two moments for the absolute area by employing \eqref{aa1m} and \eqref{aeqcA} together with the one-time and two-time PDF for the BM in heterogeneous media. In \ref{app:aam1_aam2} we detail the calculations. We find,
\begin{eqnarray}
    \left\langle \mathcal{A}(t)^{}\right\rangle =\frac{1}{\sqrt{\pi}}\frac{(2-\alpha)^{\frac{4-\alpha}{2-\alpha}}}{3-\alpha}D_{0}^{\frac{1}{2-\alpha}}\Gamma\left(\frac{4-\alpha}{4-2\alpha}\right)t^{\frac{3-\alpha}{2-\alpha}}
    \label{aam1}
\end{eqnarray}
and
\begin{eqnarray}
    \left\langle \mathcal{A}(t)^{2}\right\rangle =\frac{1}{\sqrt{\pi}}\frac{(2-\alpha)^{\frac{6-\alpha}{2-\alpha}}}{3-\alpha}\left(D_{0}/2\right)^{\frac{2}{2-\alpha}}\Gamma\left(\frac{4-\alpha}{2-\alpha}\right)\Gamma\left(\frac{4-\alpha}{2(2-\alpha)}\right)\mathcal{C}(\alpha)t^{\frac{2(3-\alpha)}{2-\alpha}}
\label{aam2}
\end{eqnarray}
where
$$
\mathcal{C}(\alpha)=\frac{\Gamma\left(\frac{10-3\alpha}{4-2\alpha}\right)}{\Gamma\left(\frac{16-5\alpha}{4-2\alpha}\right)}{}_3F_2\left(\frac{4-\alpha}{2(2-\alpha)},\frac{4-\alpha}{2(2-\alpha)},\frac{3-\alpha}{2-\alpha};\frac{1}{2},\frac{16-5\alpha}{4-2\alpha};1\right)
$$
The ergodicity breaking parameter computed from \eqref{aam1} and \eqref{aam2} is
\begin{eqnarray}
    \text{EB}_{\text{HBM}}=\mathcal{C}(\alpha)\Gamma\left(\frac{5-2\alpha}{2-\alpha}\right)-1.
    \label{EB_hbm}
\end{eqnarray}

\begin{figure}[h]
    \centering
    \includegraphics[width=0.99\linewidth]{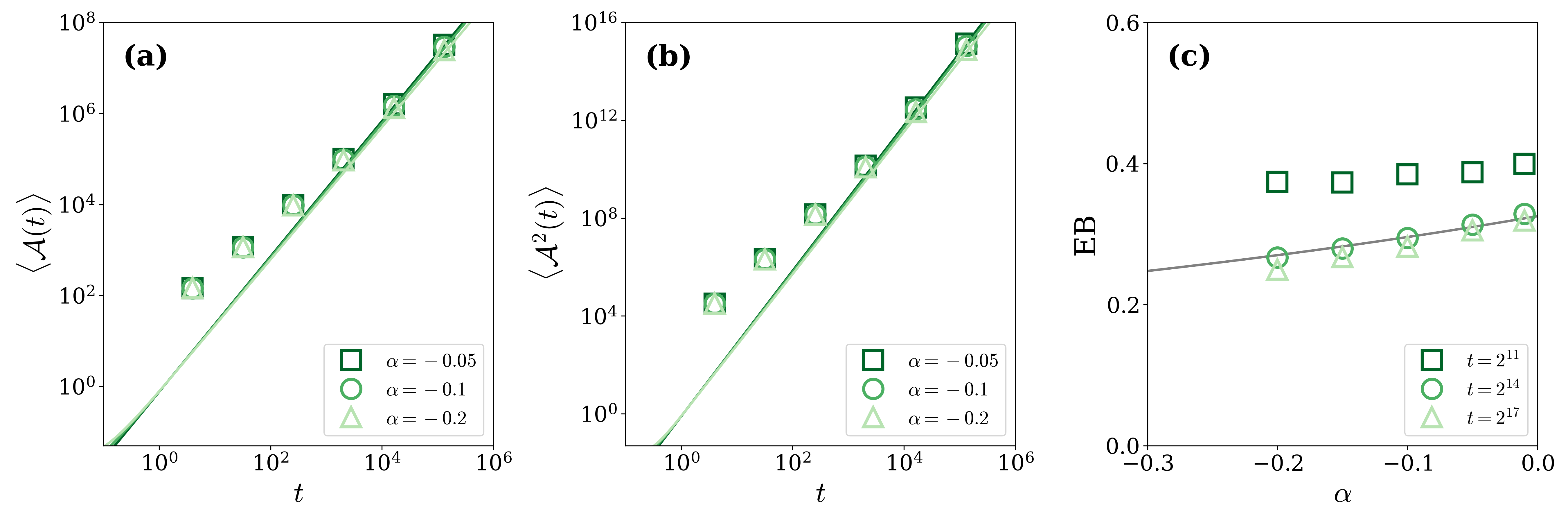}
    \caption{Heterogeneous Brownian Motion (HBM). (a) First moment of the absolute area $\langle \mathcal{A}(t) \rangle$ as a function of time $t$. The lines correspond to the theoretical prediction given by Eq.~\eqref{aa1m}. (b) Second moment of the area $\langle \mathcal{A}^{2}(t) \rangle$ as a function of time $t$. The lines correspond to the theoretical prediction given by Eq.~\eqref{aa1m}. (c) Ergodicity breaking parameter $EB$ as a function of the parameter $\alpha/H$.  The line corresponds to the theoretical prediction given by Eq.~\eqref{EB_hbm}. The simulations details are provided in \ref{app:simulations}.}
    \label{fig:HBM}
\end{figure}
In Fig. \ref{fig:HBM} we compare the above theoretical results with numerical simulations. It is observed how in this case the convergence to the long time behavior is very slow. The EB parameter increases with $\alpha$.

\subsection{EB parameter discussion}\label{sub:eb_parameter}

We now examine and compare the behavior of the EB parameter for $\mathcal{A}$ across the models presented in this work. In Fig. \ref{fig:SBM}(c) we can see that for the SBM the EB parameter decreases with $\alpha$. Since EB is a measure of the variability between trajectories regarding a certain observable this means that the lower the $\alpha$ the higher fluctuations of $\mathcal{A}$ for each trajectory. For the SBM,  since $\alpha\in(0,1)$, $D(t)$ decreases with time, making escaping from a given region progressively harder, and thus enhancing differences in $\mathcal{A}$ between trajectories. This effect is more pronounced at lower values of $\alpha$, and hence the increase in EB. We get the same result for the subdiffusive-CTRW, as plotted in Fig. \ref{fig:CTRW}(c). In this case, lower values of $\alpha$ mean larger waiting times in a given region, and therefore a higher fluctuations in the contributions to $\mathcal{A}$.

However, for the fBM we can see in Fig. \ref{fig:fBM}(c) that the EB of $\mathcal{A}$ increases with $H$. This result is in line with the previous two, since the persistence of the fBM increases with $H$, and thus EB increases too. Lastly, we look at the EB value for the HBM. In this case, see Fig. \ref{fig:HBM}(c), EB increases with $\alpha$ and vanishes at $\alpha\to-\infty$. For $\alpha<0$, $D(x)$ decreases with $x$ and is maximum at $x=0$, therefore the walker escapes fast from the origin and as $x$ increases gets progressively slowed, acting as an effective binding potential. Indeed, the PDF $P(x,t)$ in the subdiffusive regime, $\alpha<0$, is zero at $x=0$ and decays in space as $P(x,t)\sim |x|^{|\alpha|/2}\exp\left(-A|x|^{2+|\alpha|}/t\right)$, where $A>0$ is a parameter (see Eq. \eqref{pdfdx}). Thus, as $|x|\to\infty$, the diffusion coefficient vanishes and trajectories become effectively trapped at the system boundaries. The more negative $\alpha$ is, the fewer steps are required to reach this limiting regime. Since we consider the absolute value of the area, left- and right-moving trajectories contribute identically to the observable. Consequently, for increasingly negative $\alpha$, trajectories need to explore less space before reaching this boundary, leading to very similar absolute areas across realizations and, therefore, to a reduction of EB.

Finally, it is worth comparing these results to the  EB parameter previously obtained for other functionals of the random walk, for example, the half occupation time $T^+$, which is the time that the walker spends on the positive half-space $x>0$. We have studied this functional in  \cite{MeFlPa25} for the subdiffusive-CTRW, in \cite{MeHeFl25} for the fBM and SBM, and in \cite{MeFl26} for the HBM. Remarkably, we find that for the SBM, fBM, and subdiffusive-CTRW, the EB of $T^+$ have the same qualitative behavior with $\alpha$ (or $H$) as for the EB of $\mathcal{A}$. However, this is not the case for the HBM: the EB parameter for $T^+$ increases with $\alpha$, instead of decreasing. The key difference is that, in the case of spatial heterogeneity, all trajectories are confined to the same spatial range, whereas in the other processes, the differences between trajectories are of temporal origin. For example, in SBM the diffusion coefficient vanishes, but at spatial positions that differ from trajectory to trajectory; the resulting immobilization thus increases EB, as the locations where trapping occurs, and hence the contributions to the area and occupation times, are highly heterogeneous. A similar argument applies to subdiffusive-CTRW and fBM. This comparison highlights the role of the functional of study, and how the same underlying random walk gives rise to different ergodic behaviors depending on the observable.

\section{PDF}\label{sec:PDFs}

In Sec. \ref{sec:area} and Sec. \ref{sec:absolute_area}, we have derived the first two moments of $A$ and $\mathcal{A}$ for two Gaussian processes, the SBM and the fBM, and two non-Gaussian processes, a subdiffusive-CTRW and a HBM. In this section we will study the general properties of the PDFs of $A$ and $\mathcal{A}$ for all these models. To do so we begin by connecting the MSD of the random walk with the second moment of $A$ and $\mathcal{A}$. Suppose that $\gamma$ is the exponent of the MSD:  $\left\langle x(t)^{2}\right\rangle\sim t^\gamma$. Rewriting the second moment of the area (or absolute area), for all processes studied in this work, in terms of their corresponding $\gamma$ we find that the scaling is as 
\begin{align}
    \left\langle Z(t)^{2}\right\rangle\sim t^{2+\gamma},
    \label{eq:z2_scaling}
\end{align}
where $Z$ is either $A$ or $\mathcal{A}$. To understand this result, recall that by the definition of $\gamma$ we have $x\sim t^{\gamma/2}$, and then $Z\sim\int^t \tau^{\gamma/2}d\tau\sim t^{1+\gamma/2}$, yielding the $t^{2+\gamma}$ scaling for the second moment. Note that the second moment of both the area and the absolute area have the same temporal scaling. In the case of the absolute area, we can follow the same reasoning to find that the scaling of the first moment goes as $\langle \mathcal{A}(t) \rangle \sim t^{1+\gamma/2}$. Then, we see that for the absolute are the first two moments obey the simple scaling $\left\langle \mathcal{A}(t)^2\right\rangle \sim \left\langle \mathcal{A}(t)\right\rangle ^2 $.

Once we have found a general scaling for the moments of $A$ and $\mathcal{A}$, we can compute the scaling of the $P(Z,t)$. Assume that in the long time limit the PDF of $Z(t)$ has the scaling form  $P(Z,t)\sim t^{-\eta}g(Z/t^{\rho})$ where $g(\cdot)$ is the scaling function and the exponents $\eta$ and $\rho$ have to be found. Since the integral $\int_{-\infty}^{\infty}P(Z,t)dZ=t^{\rho-\eta}\int_{-\infty}^{\infty}h(u)du$ must be time-independent (and equal to 1) both exponents must be equal, and therefore $\rho=\eta$. We can determine the remaining exponent through the scaling of the second moment, which we know. The second moment is $\left\langle Z(t)^{2}\right\rangle =\int_{-\infty}^{\infty}Z^{2}P(Z,t)dZ\sim t^{2\eta}\int_{-\infty}^{\infty}u^{2}h(u)du$. Comparing this scaling with \eqref{eq:z2_scaling} we have $\eta=1+\gamma/2$. Thus, we have obtained a general scaling for both the PDF of $A$ and $\mathcal{A}$,
\begin{align}
P(Z,t)\sim \frac{1}{t^{1+\frac{\gamma}{2}}}g_\gamma\left(\frac{Z}{t^{1+\frac{\gamma}{2}}}\right)
\label{eq:scaling}
\end{align}
in terms of the scaling exponent of the MSD of the random walk $\gamma$. Note that $g_\gamma(\cdot)$ is different for the different models as well as observables, and, in general, depends on $\gamma$. The specific values of $\gamma$ for the models discussed in this work are: $\gamma_{BM}=1$, $\gamma_{SMB}=\alpha$, $\gamma_{fBM}=2H$, $\gamma_{CTRW}=\alpha$, and $\gamma_{HBM}=2/(2-\alpha)$. 

For the area $A$ under a Gaussian process, we can go beyond the scaling and derive a closed form expression for $P(A,t)$. If $x(t)$ follows a Gaussian process with zero
mean and autocorrelation $C(t_1,t_2)$ , the characteristic functional of $x(t)$ is given by \cite{vK92} and is written in Eq. \eqref{cf}. On the other hand, the characteristic function of $A(t)$ is its Fourier transform
\begin{eqnarray}
P(k,t)=\mathcal{F}_{Z\to k}[P(A,t)]=\int_{-\infty}^{\infty}e^{ikA}P(A,t)dA=\left\langle e^{ikA(t)}\right\rangle .
\label{cf2}
\end{eqnarray}
Eq. \eqref{cf2} can be linked to the characteristic functional of $x(t)$ by considering $f(\tau)=k\mathds{1}_{[0,t]}(\tau)$ in \eqref{cf}, where $\mathds{1}_{[a,b]}(x)$ is the indicator function,
\begin{eqnarray}
    P(k,t)=\left\langle e^{ikA(t)}\right\rangle =\left\langle e^{ik\int_{0}^{t}x(\tau)d\tau}\right\rangle =e^{-\frac{k^{2}}{2}\left\langle A(t)^{2}\right\rangle }.
    \label{mgfA}
\end{eqnarray}
Performing the inverse Fourier transform we finally find
\begin{eqnarray}
    P(A,t)=\frac{1}{\sqrt{2\pi\left\langle A(t)^{2}\right\rangle }}e^{-\frac{A^{2}}{2\left\langle A(t)^{2}\right\rangle }}.
    \label{gauss}
\end{eqnarray}
This result is expected: since $A(t)$ is the sum of Gaussian processes $x(t)$ then $A(t)$ is also a Gaussian process. In consequence, the PDF of $A$ must be a Gaussian distribution. Notice that indeed Eq. \eqref{gauss} follows the scaling derived in Eq. \eqref{eq:scaling}. For the models discussed in this work, we get
\begin{eqnarray}
    P(A,t)_{\textrm{BM}}&=&\frac{1}{\sqrt{4\pi Dt^{3}/3}}e^{-\frac{3A^{2}}{4Dt^{3}}}, \label{pabm} \\
    P(A,t)_{\textrm{SBM}}&=&\sqrt{\frac{(\alpha+1)(\alpha+2)}{8\pi Kt^{\alpha+2}}}\exp\left[-\frac{(\alpha+1)(\alpha+2)}{8Kt^{\alpha+2}}A^{2}\right], \label{pdf_sbm} \\
    P(A,t)_{\textrm{fBM}}&=&\sqrt{\frac{H+1}{2\pi D_{H}t^{2H+2}}}\exp\left[-\frac{(H+1)A^{2}}{2D_{H}t^{2H+2}}\right]. \label{pdf_fbm}
\end{eqnarray}

\begin{figure}[h]
    \centering
    \includegraphics[width=0.99\linewidth]{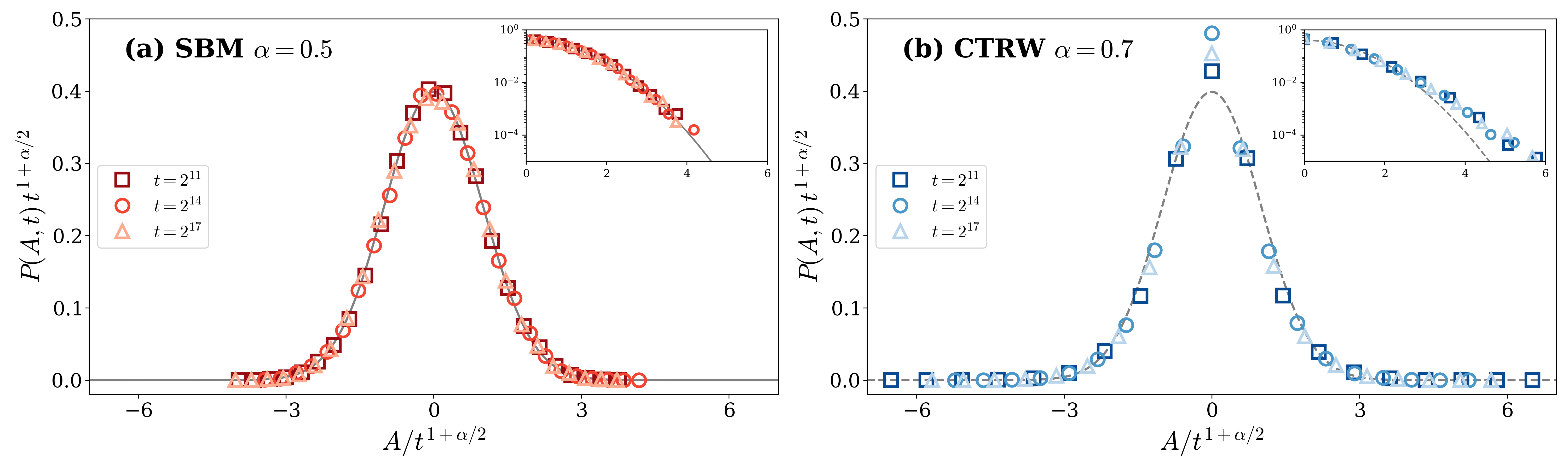}
    \caption{ Probability distribution of $P(A,t)$ vs $A$ at different times $t$ . (a) Scaled Brownian motion (SBM). The line corresponds to the theoretical prediction given by Eq.~\eqref{pdf_sbm}.  (b) Continuous Time Random Walk (CTRW). The dotted line corresponds to the prediction given if the functional was Gaussian. The simulations details are provided in \ref{app:simulations}.}
    \label{fig:pdf_A_main}
\end{figure}

\begin{figure}[h]
    \centering
    \includegraphics[width=0.99\linewidth]{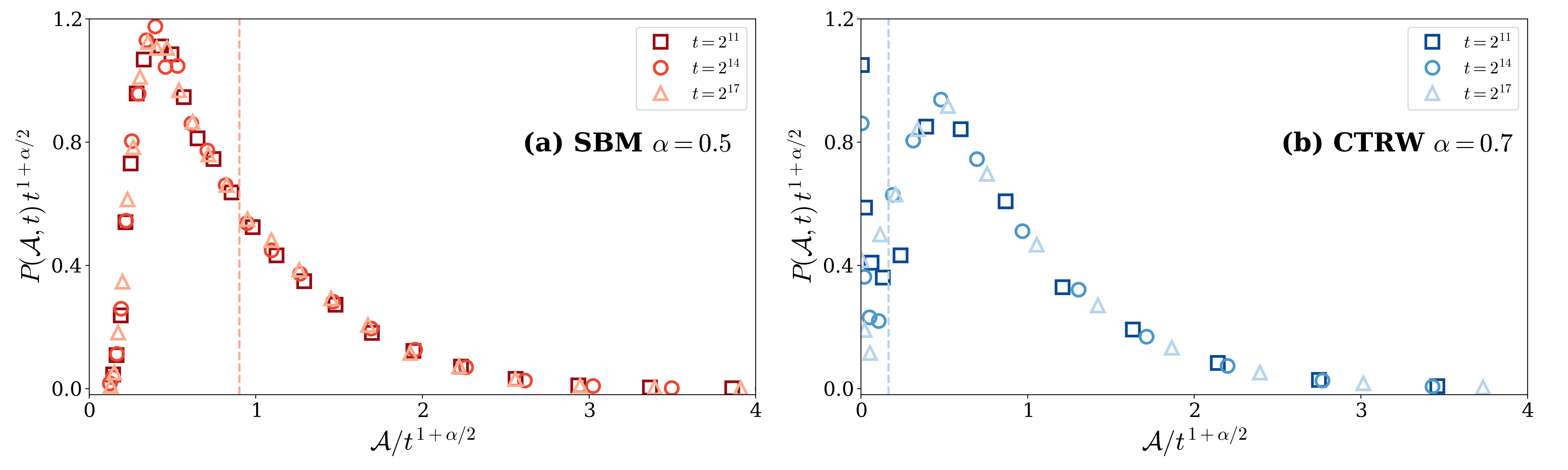}
    \caption{ Probability distribution of $P(\mathcal{A,t})$ vs $\mathcal{A}$ at different times $t$ . (a) Fractional Brownian Motion (fBM). The vertical line corresponds to the mean value predicted by Eq. \eqref{aiaaaa}. (b) Heterogeneous Brownian motion (HBM). The vertical line corresponds to the mean value predicted by Eq. \eqref{aa12}. The simulations details are provided in \ref{app:simulations}. }
   \label{fig:pdf_Aabs_main}
\end{figure}

The results for the area under a BM connect to the random acceleration process (when the particle's velocity is a Wiener process or the acceleration is a white Gaussian noise) \cite{Ch43}, where the particle position follows the same distribution as \eqref{pabm}. In Fig. \ref{fig:pdf_A_main} we study the scaling of the $P(A,t)$ at different simulation times for both a Gaussian -SBM in panel a)-, and a non-Gaussian model -CTRW in panel b). In both cases, we see that the scaling derived in Eq. \eqref{eq:scaling} is satisfied. Moreover, in both panels we compare the data to a corresponding Gaussian PDF, Eq. \eqref{gauss}. For the SBM, we see that the data indeed follows a Gaussian distribution, as discussed above. This is not the case for the CTRW. In this case, we can see that the probability around $A=0$ is larger than the corresponding to a Gaussian PDF, and that the decay is slower at the tails, as can be appreciated in the inset, this deviation is in agreement with \cite{BaBu20}. Finally, in Fig. \ref{fig:pdf_Aabs_main} we plot the PDF for the absolute area $\mathcal{A}$, and again we can see that the scaling predicted by Eq. \eqref{eq:scaling} is fulfilled. The peak near zero for the CTRW -panel b)- comes from the trajectories that at the end of the simulation time have not jumped for the first time yet, and is in agreement with the predicted value of $\langle\mathcal{A}(t)\rangle$. In \ref{app:figures} we include the same analysis for the fBM and HBM.

\section{Conclusions}\label{sec:conclusions}
In this paper, we investigate the statistical properties of the area and absolute area under trajectories of subdiffusive random walks considering different models such as the scaled Brownian motion, fractional Brownian motion, subdiffusive-continuous-time random walks, and Brownian motion in heterogeneous media. We compute the first two moments for both functionals, the ergodicity-breaking parameter for the absolute area, and the scaling forms of the associated probability density functions for both functionals. Although the temporal scaling of these functionals is universal across the models considered, the prefactors of the moments, the behavior of the ergodicity breaking parameter with the anomalous exponents, and the nature of the tails of the probability density function differ, allowing one to discriminate between the underlying mechanisms of subdiffusion. These findings suggest that measurements of such functionals, for example in NMR experiments, can provide insight into the microscopic origin of subdiffusive motion of spin-bearing particles.

%The measure of the signal in NMR experiments can be useful to uncover the microscopic origin of subdiffusive motion of particles with spin. Although the temporal dependence for the area and the absolute area is the same across the different microscopic approaches to subdiffusion, the prefactors of the moments are different and this can allow to discriminate between the different approaches by comparing the experimental results with the theoretical predictions. Moreover, the analysis of the tails of $P(A,t)$, and in particular if they decay as Gaussian, faster-than-Gaussian or slower-than-Gaussian, can further help distinguish between the different subdiffusive models.

\section*{Acknowledgements}
The authors acknowledge the financial support of the Ministerio de Ciencia e Innovaci\'on (Spanish government) under
Grant No. PID2021-122893NB-C22.

\appendix
\section{Derivation of Eq. \eqref{a2sct} } \label{app:a2sct}
Let us start from the autocorrelation function of the area
\begin{eqnarray}
    \left\langle A(t_{1})A(t_{2})\right\rangle =\int_{0}^{t_{1}}d\tau_{1}\int_{0}^{t_{2}}d\tau_{2}\left\langle x(\tau_{1})x(\tau_{2})\right\rangle .
    \label{aca}
\end{eqnarray}
We introduce the double Laplace transform with respect to $t_1$ and $t_2$ with Laplace variables $\lambda_1$ and $\lambda_2$ as 
$$
\left\langle \tilde{A}(\lambda_{1})\tilde{A}(\lambda_{2})\right\rangle =\mathcal{L}_{t_{1}\to\lambda_{1}}\mathcal{L}_{t_{2}\to\lambda_{2}}\left\langle A(t_{1})A(t_{2})\right\rangle =\int_{0}^{\infty}e^{-\lambda_{1}t_{1}}dt_{1}\int_{0}^{\infty}e^{-\lambda_{2}t_{2}}dt_{2}\left\langle A(t_{1})A(t_{2})\right\rangle .
$$
Transforming \eqref{aca} one finds
\begin{eqnarray}
    \left\langle \tilde{A}(\lambda_{1})\tilde{A}(\lambda_{2})\right\rangle =\frac{1}{\lambda_{1}\lambda_{2}}\left\langle \tilde{x}(\lambda_{1})\tilde{x}(\lambda_{2})\right\rangle 
    \label{AAl}
\end{eqnarray}
where the tilde symbol means Laplace transformed function. Here, the trajectory $\{x(\tau); 0\leq \tau\leq t\}$ is the corresponding to a particle moving according to a subdiffusive-CTRW. To find the autocorrelation in the Laplace space $\left\langle \tilde{x}(\lambda_{1})\tilde{x}(\lambda_{2})\right\rangle$ in terms of the autocorrelation corresponding to a Brownian particle we make use of the subordination between both processes \cite{Ca15}:
\begin{eqnarray}
    \left\langle \tilde{x}(\lambda_{1})\tilde{x}(\lambda_{2})\right\rangle =\int_{0}^{\infty}ds_{1}\int_{0}^{\infty}ds_{2}\left\langle x(s_{1})x(s_{2})\right\rangle _{\textrm{BM}}h(s_{2},\lambda_{2},s_{1},\lambda_{1} )
    \label{cova}
\end{eqnarray}
where $h(s_{2},\lambda_{2},s_{1},\lambda_{1} )$ is the two-point PDF of subordination and is given by \cite{Ca15}
\begin{eqnarray}
    h(s_{2},\lambda_{2},s_{1},\lambda_{1})&=&M(\lambda_{2},\lambda_{1})e^{-s_{1}\Phi(\lambda_{1}+\lambda_{2})}\delta(s_{2}-s_{1})\nonumber\\
    &+&L_{1}(\lambda_{2},\lambda_{1})e^{-s_{1}\Phi(\lambda_{1}+\lambda_{2})-(s_{2}-s_{1})\Phi(\lambda_{2})}\theta(s_{2}-s_{1})\nonumber\\
    &+&L_{2}(\lambda_{2},\lambda_{1})e^{-s_{2}\Phi(\lambda_{1}+\lambda_{2})-(s_{1}-s_{2})\Phi(\lambda_{1})}\theta(s_{1}-s_{2})
    \label{h}
\end{eqnarray}
where $\delta(\cdot)$ and $\theta(\cdot)$ are the Dirac delta and Heaviside step functions respectively, and 
\begin{eqnarray*}
    M(\lambda_{2},\lambda_{1})&=&\frac{\Phi(\lambda_{1})+\Phi(\lambda_{2})-\Phi(\lambda_{1}+\lambda_{2})}{\lambda_{1}\lambda_{2}}\nonumber\\
    L_{1}(\lambda_{2},\lambda_{1})&=&\frac{\Phi(\lambda_{2})\left[\Phi(\lambda_{1}+\lambda_{2})-\Phi(\lambda_{2})\right]}{\lambda_{1}\lambda_{2}}\nonumber\\
    L_{2}(\lambda_{2},\lambda_{1})&=&\frac{\Phi(\lambda_{2})\left[\Phi(\lambda_{1}+\lambda_{2})-\Phi(\lambda_{1})\right]}{\lambda_{1}\lambda_{2}}.
\end{eqnarray*}
The function $\Phi (\lambda)$ is related to the waiting time PDF in the Laplace space and when the later is given by \eqref{ml} then the former is $\Phi(\lambda)=\lambda^\alpha \tau^{\alpha -1}$ \cite{Ca15}. Taking into account that $\left\langle x(s_{1})x(s_{2})\right\rangle _{\textrm{BM}}=2D\min (s_1,s_2)$ the integrals in Eq. \eqref{cova} can be straightforwardly calculated and from Eq. \eqref{AAl}  we get
$$
\left\langle \tilde{A}(\lambda_{1})\tilde{A}(\lambda_{2})\right\rangle =\frac{2D}{\tau^{\alpha-1}\lambda_{1}^2\lambda_{2}^2(\lambda_{1}+\lambda_{2})^{\alpha}}.
$$
The task now is to perform the double Laplace inversion to this correlation. We perform first the inverse Laplace transform with respect to $\lambda_2$ considering $\mathcal{L}_{\lambda_{2}\to t_{2}}^{-1}\left[\lambda_{2}^{-2}\right]=t_{2}$, using the shifting property
$$
\mathcal{L}_{\lambda_{2}\to t_{2}}^{-1}\left[\frac{1}{\left(\lambda_{1}+\lambda_{2}\right)^{\alpha}}\right]=\frac{e^{-\lambda_{1}t_{2}}}{t_{2}^{1-\alpha}\Gamma(\alpha)},
$$
and the convolution
$$
\mathcal{L}_{\lambda_{2}\to t_{2}}^{-1}\left[\frac{1}{\left(\lambda_{1}+\lambda_{2}\right)^{\alpha}\lambda_{2}^{2}}\right]=\frac{1}{\Gamma(\alpha)}\int_{0}^{t_{2}}d\tau_{2}\frac{e^{-\lambda_{1}\tau_{2}}}{\tau_{2}^{1-\alpha}}(t_{2}-\tau_{2})
$$
we get
$$
\left\langle \tilde{A}(\lambda_{1})A(t_{2})\right\rangle =\mathcal{L}_{\lambda_{2}\to t_{2}}^{-1}\left[\left\langle \tilde{A}(\lambda_{1})\tilde{A}(\lambda_{2})\right\rangle \right]=\frac{2D}{\tau^{\alpha-1}\lambda_{1}^{2}\Gamma(\alpha)}\int_{0}^{t_{2}}d\tau_{2}\frac{e^{-\lambda_{1}\tau_{2}}}{\tau_{2}^{1-\alpha}}(t_{2}-\tau_{2}).
$$
Inverting now with respect to $t_1$,
\begin{eqnarray*}
   \left\langle A(t_{1})A(t_{2})\right\rangle &=&\mathcal{L}_{\lambda_{1}\to t_{1}}^{-1}\left[\left\langle \tilde{A}(\lambda_{1})A(t_{2})\right\rangle \right]=\frac{2D}{\tau^{\alpha-1}\Gamma(\alpha)}\int_{0}^{t_{2}}d\tau_{2}\frac{t_{2}-\tau_{2}}{\tau_{2}^{1-\alpha}}\mathcal{L}_{\lambda_{1}\to t_{1}}^{-1}\left[\frac{e^{-\lambda_{1}\tau_{2}}}{\lambda_{1}^{2}}\right]\nonumber\\
   &=&\frac{2D}{\tau^{\alpha-1}\Gamma(\alpha)}\int_{0}^{t_{2}}d\tau_{2}\frac{(t_{2}-\tau_{2})(t_{1}-\tau_{2})}{\tau_{2}^{1-\alpha}}\theta(t_{1}-\tau_{2}). 
\end{eqnarray*}
Finally, performing the integral we obtain the auto-correlation of the area
$$
\left\langle A(t_{1})A(t_{2})\right\rangle =\frac{2D\min(t_{1},t_{2})^{\alpha}}{\tau^{\alpha-1}\Gamma(\alpha)}\left[\frac{t_{1}t_{2}}{\alpha}-\frac{t_{1}+t_{2}}{\alpha+1}\min(t_{1},t_{2})+\frac{\min(t_{1},t_{2})^{2}}{\alpha+2}\right].
$$
Replacing $t_1=t_2=t$ we finally find Eq. \eqref{a2sct}.

\section{Derivation of Eq. \eqref{A2dx} } \label{app:A2dx}
To compute $\left\langle A(t)^{2}\right\rangle $ we need the two-time PDF $P(x_2,t_2;x_1,t_1)$ but it is very difficult to obtain from \eqref{led2}. However, performing the transformation in Eq. \eqref{y(x)} we can integrate over the variable $y$. Hence, from \eqref{eqcA}
\begin{eqnarray}
    \left\langle A(t)^{2}\right\rangle &=&2\int_{0}^{t}d\tau_{2}\int_{0}^{\tau_{2}}d\tau_{1}\int_{-\infty}^{\infty}x_{2}dx_{2}\int_{-\infty}^{\infty}x_{1}dx_{1}P(x_{2},\tau_{2}|x_{1},\tau_{1})P(x_{1},\tau_{1})\nonumber\\
    &=&2\int_{0}^{t}d\tau_{2}\int_{0}^{\tau_{2}}d\tau_{1}\int_{-\infty}^{\infty}x_{2}(y_{2})dy_{2}\int_{-\infty}^{\infty}x_{1}(y_{1})dy_{1}P(y_{2},\tau_{2}|y_{1},\tau_{1})P(y_{1},\tau_{1}).
\end{eqnarray}
Inverting \eqref{y(x)} to get $x(y)$ we find
\begin{eqnarray}
    \left\langle A(t)^{2}\right\rangle &=&2\left(\frac{2-\alpha}{\sqrt{2}}\sqrt{D_{0}}\right)^{\frac{4}{2-\alpha}}\int_{0}^{t}d\tau_{2}\int_{0}^{\tau_{2}}d\tau_{1}\int_{-\infty}^{\infty}|y_{2}|^{\frac{2}{2-\alpha}}\textrm{sign}(y_{2})dy_{2}\nonumber\\
    &\times &\int_{-\infty}^{\infty}|y_{1}|^{\frac{2}{2-\alpha}}\textrm{sign}(y_{1})dy_{1}P(y_{2},\tau_{2}|y_{1},\tau_{1})P(y_{1},\tau_{1}).
    \label{A2ens}
\end{eqnarray}
From \eqref{oty} and the two-time PDF of $y(t)$
$$
P(y_{2},\tau_{2}|y_{1},\tau_{1})=\frac{1}{\sqrt{2\pi(\tau_{2}-\tau_{1})}}\exp\left[-\frac{(y_{2}-y_{1})^{2}}{2(\tau_{2}-\tau_{1})}\right]
$$
we can perform the integration over $y_2$ in Eq. \eqref{A2ens} to get
\begin{eqnarray*}
   & &\int_{-\infty}^{\infty}|y_{2}|^{\frac{2}{2-\alpha}}\textrm{sign}(y_{2})P(y_{2},\tau_{2}|y_{1},\tau_{1})dy_{2}\\
   &=&\frac{1}{\sqrt{2\pi(\tau_{2}-\tau_{1})}}\left\{ \int_{0}^{\infty}z^{\frac{2}{2-\alpha}}\exp\left[-\frac{(z-y_{1})^{2}}{2(\tau_{2}-\tau_{1})}\right]dz-\int_{0}^{\infty}z^{\frac{2}{2-\alpha}}\exp\left[-\frac{(z+y_{1})^{2}}{2(\tau_{2}-\tau_{1})}\right]dz\right\} 
\end{eqnarray*}
Each of the above integrals can be expressed in terms of parabolic cylinder functions $D_\nu (z)$ using Eq. 3.462.1 of Ref. \cite{Gr15}. Therefore,
\begin{eqnarray*}
    & &\int_{-\infty}^{\infty}|y_{2}|^{\frac{2}{2-\alpha}}\textrm{sign}(y_{2})P(y_{2},\tau_{2}|y_{1},\tau_{1})dy_{2}\\
   &=&\frac{\left(\tau_{2}-\tau_{1}\right)^{\frac{1}{2-\alpha}}}{\sqrt{2\pi}}\Gamma\left(\frac{4-\alpha}{2-\alpha}\right)e^{-\frac{y_{1}^{2}}{4(\tau_{2}-\tau_{1})}}\left[D_{-\frac{4-\alpha}{2-\alpha}}\left(-\frac{y_{1}}{\sqrt{\tau_{2}-\tau_{1}}}\right)-D_{-\frac{4-\alpha}{2-\alpha}}\left(\frac{y_{1}}{\sqrt{\tau_{2}-\tau_{1}}}\right)\right].
\end{eqnarray*}
On the other hand, parabolic cylinder functions can be related to confluent hypergeometric functions using the property (see Eq. 9.240 of Ref. \cite{Gr15} )
$$
D_{-\nu}(-z)-D_{-\nu}(z)=\frac{\sqrt{\pi}2^{\frac{3-\nu}{2}}}{\Gamma(\nu/2)}ze^{-\frac{z^{2}}{4}}M\left(\frac{\nu+1}{2},\frac{3}{2},\frac{z^{2}}{2}\right)
$$
$$
\frac{2^{\frac{3-\alpha}{2-\alpha}}}{\sqrt{2\pi}}\Gamma\left(\frac{3-\alpha}{2-\alpha}\right)\left(\tau_{2}-\tau_{1}\right)^{\frac{\alpha}{2(2-\alpha)}}y_{1}e^{-\frac{y_{1}^{2}}{2(\tau_{2}-\tau_{1})}}M\left(\frac{3-\alpha}{2-\alpha},\frac{3}{3},\frac{y_{1}^{2}}{2(\tau_{2}-\tau_{1})}\right)
$$
where where $M(a,b,z)$ is the Kummer $M$ function (see Chapter 13 in Ref. \cite{ab64}). Now, performing the integration over $y_1$ using Eq. 7.621.4 of Ref. \cite{Gr15} we find
\begin{eqnarray*}
    & &\int_{-\infty}^{\infty}|y_{2}|^{\frac{2}{2-\alpha}}\textrm{sign}(y_{2})dy_{2}\int_{-\infty}^{\infty}|y_{1}|^{\frac{2}{2-\alpha}}\textrm{sign}(y_{1})dy_{1}P(y_{2},\tau_{2}|y_{1},\tau_{1})P(y_{1},\tau_{1})\\
    &=&\frac{2^{\frac{4-\alpha}{2-\alpha}}}{\pi}\Gamma\left(\frac{3-\alpha}{2-\alpha}\right)^{2}\left(1-\frac{\tau_{1}}{\tau_{2}}\right)^{\frac{6-\alpha}{2(2-\alpha)}}\sqrt{\frac{\tau_{1}}{\tau_{2}}}\left(\tau_{1}\tau_{2}\right)^{\frac{1}{2-\alpha}}{}_{2}F_{1}\left(\frac{3-\alpha}{2-\alpha},\frac{3-\alpha}{2-\alpha};\frac{3}{2},\frac{\tau_{1}}{\tau_{2}}\right)
\end{eqnarray*}
where ${}_{2}F_{1}(\cdot)$ is the hypergeometric function. Next, we have to integrate over $\tau_2$ and $\tau_1$. To do this we introduce the new variable $u=\tau_1/\tau_2$ to integrate first over $\tau_1$ and find

\begin{eqnarray}
    \left\langle A(t)^{2}\right\rangle =\frac{2}{\pi}\frac{(2-\alpha)^{\frac{6-\alpha}{2-\alpha}}}{3-\alpha}D_{0}^{\frac{2}{2-\alpha}}\Gamma\left(\frac{3-\alpha}{2-\alpha}\right)^{2}\mathcal{B}(\alpha)t^{\frac{2(3-\alpha)}{2-\alpha}}
\end{eqnarray}
with
\begin{eqnarray*}
  \mathcal{B}(\alpha)&=&\int_{0}^{1}u^{\frac{4-\alpha}{2(2-\alpha)}}(1-u)^{\frac{6-\alpha}{2(2-\alpha)}}{}_{2}F_{1}\left(\frac{3-\alpha}{2-\alpha},\frac{3-\alpha}{2-\alpha};\frac{3}{2},u\right)du\\
  &=&\frac{\Gamma\left(\frac{8-3\alpha}{4-2\alpha}\right)\Gamma\left(\frac{10-3\alpha}{4-2\alpha}\right)}{\Gamma\left(\frac{9-3\alpha}{2-\alpha}\right)}{}_3F_2\left(\frac{3-\alpha}{2-\alpha},\frac{3-\alpha}{2-\alpha},\frac{8-3\alpha}{4-2\alpha};\frac{3}{2},\frac{9-3\alpha}{2-\alpha};1\right).  
\end{eqnarray*}
where we have used Eq. 7.512.5 of Ref. \cite{Gr15}.
 
\section{Derivation of Eq. \eqref{aiaaaa}} \label{app:aiaaaa}
The first moment can be readily found inserting \eqref{cmsd} into \eqref{aa1}. We detail here the derivation of the second moment for the absolute area.  Inserting the autocorrelation \eqref{cmsd} into Eq. \eqref{aa2} and introducing the new variables $u=\tau_1/\tau_2$ and $v=\tau_2/t$ we write
$$
\left\langle \mathcal{A}(t)^2\right\rangle =\frac{4K}{\pi}t^{\alpha+2}\int_{0}^{1}v^{\alpha+1}dv\left[\frac{\pi}{\alpha+1}-2\int_{0}^{1}u^{\alpha}\arctan\left(\sqrt{\frac{1}{u^{\alpha}}-1}\right)du+2\int_{0}^{1}u^{\alpha/2}\sqrt{1-u^{\alpha}}du\right].
$$
The integral over $v$ can be easily performed. The first integral inside the parenthesis can be evaluated as follows. First we introduce the new variable $y=\sqrt{u^{-\alpha}-1}$ to get
$$
\int_{0}^{1}u^{\alpha}\arctan\left(\sqrt{\frac{1}{u^{\alpha}}-1}\right)du=\frac{2}{\alpha}\int_{0}^{\infty}\frac{y\arctan(y)}{\left(1+y^{2}\right)^{2+1/\alpha}}dy,
$$
now integrating by parts
$$
\frac{2}{\alpha}\int_{0}^{\infty}\frac{y\arctan(y)}{\left(1+y^{2}\right)^{2+1/\alpha}}dy=\frac{1}{\alpha+1}\int_{0}^{\infty}\frac{dy}{\left(1+y^{2}\right)^{2+1/\alpha}}.
$$
Making use of the definition of the beta function (see Eq. 3.194.3 in Ref. \cite{Gr15}) 
$$
\int_{0}^{1}\frac{y^{\mu-1}}{(1+y)^{\nu}}dy=B(\mu,\nu-\mu)=\frac{\Gamma(\mu)\Gamma(\nu-\mu)}{\Gamma(\nu)},\quad\nu>\mu>0.
$$
Again, using this integral we find for the second integral inside parenthesis
$$
\int_{0}^{1}u^{\alpha/2}\sqrt{1-u^{\alpha}}du=\frac{\sqrt{\pi}}{2(\alpha+1)}\frac{\Gamma\left(\frac{1}{2}+\frac{1}{\alpha}\right)}{\Gamma\left(\frac{1}{4}+\frac{1}{\alpha}\right)}.
$$
Finally, using the property $\Gamma(1+z)=z\Gamma(z)$ and simplifying we get Eq. \eqref{aiaaaa}. 

\section{Derivation of Eq. \eqref{aa12}} \label{app:aa12}
The Montroll-Weiss equation of the CTRW is the expression for the one-time PDF of $x(t)$ in the Fourier-Laplace space. For subdiffusion in the continuum limit it is given by Eq. 31 in Ref. \cite{MeKl01}). In the Laplace space it reads
$$
P(x,s)=\frac{1}{\sqrt{2K_{\alpha}s^{2-\alpha}}}e^{-\frac{|x|s^{\alpha/2}}{\sqrt{K_{\alpha}}}}.
$$
On the other hand, Laplace transforming Eq. \eqref{aa1} we have
$$
\left\langle \mathcal{A}(s)\right\rangle =\mathcal{L}_{t\to s}\left[\left\langle \mathcal{A}(t)\right\rangle \right]=\frac{2}{s}\int_{0}^{\infty}xP(x,s)dx=\frac{\sqrt{K_{\alpha}}}{s^{2+\frac{\alpha}{2}}}
$$
which after Laplace inversion yields the first equation in Eq. \eqref{aa12}. We compute the second moment making use of subordination as in \ref{app:a2sct}. Performing the double Laplace transform of the autocorrelation function of the absolute area
\begin{eqnarray}
    \left\langle \mathcal{A}(t_{1})\mathcal{A}(t_{2})\right\rangle =\int_{0}^{t_{1}}d\tau_{1}\int_{0}^{t_{2}}d\tau_{2}\left\langle |x(\tau_{1})||x(\tau_{2})|\right\rangle 
\end{eqnarray}
we find
\begin{eqnarray}
    \left\langle \tilde{\mathcal{A}}(\lambda_{1})\tilde{\mathcal{A}}(\lambda_{2})\right\rangle =\frac{1}{\lambda_{1}\lambda_{2}}\left\langle |\tilde{x}(\lambda_{1})||\tilde{x}(\lambda_{2})|\right\rangle .
    \label{alal}
\end{eqnarray}
On the other hand, the autocorrelation of the subdiffusive-CTRW in the double Laplace space $\left\langle |\tilde{x}(\lambda_{1})||\tilde{x}(\lambda_{2})|\right\rangle$ is found from Eq. \eqref{cova} as
\begin{eqnarray}
    \left\langle |\tilde{x}(\lambda_{1})||\tilde{x}(\lambda_{2})|\right\rangle =\int_{0}^{\infty}ds_{1}\int_{0}^{\infty}ds_{2}\left\langle |x(s_{1})||x(s_{2})|\right\rangle _{\textrm{BM}}h(s_{2},\lambda_{2},s_{1},\lambda_{1})
    \label{xxl}
\end{eqnarray}
in terms of the autocorrelation of the absolute position of  BM in real time space $\left\langle |x(s_{1})||x(s_{2})|\right\rangle _{\textrm{BM}}$ which is given by
\begin{eqnarray}
   & &\left\langle |x(s_{1})||x(s_{2})|\right\rangle _{\textrm{BM}}=4\int_{0}^{\infty}x_{2}dx_{2}\int_{0}^{\infty}x_{1}dx_{1}P(x_{2},s_{2};x_{1},s_{1})\nonumber\\
    &=&\frac{2D}{\pi}\min(s_{1},s_{2})\left[\pi-2\arctan\left(\sqrt{\frac{s_{1}s_{2}}{\min(s_{1},s_{2})^{2}}-1}\right)+2\sqrt{\frac{s_{1}s_{2}}{\min(s_{1},s_{2})^{2}}-1}\right]
    \label{xx}
\end{eqnarray}
where the two-time PDF is given by \eqref{cftt}. Plugging \eqref{xx} into \eqref{xxl} using \eqref{h} we find
\begin{eqnarray}
    \left\langle |\tilde{x}(\lambda_{1})||\tilde{x}(\lambda_{2})|\right\rangle &=&2DM(\lambda_{2},\lambda_{1})\int_{0}^{\infty}e^{-s_{1}\Phi(\lambda_{2}+\lambda_{1})}s_{1}ds_{1}\nonumber\\
    &+&L_{1}(\lambda_{2},\lambda_{1})\int_{0}^{\infty}e^{-s_{1}\Phi(\lambda_{2}+\lambda_{1})+s_{1}\Phi(\lambda_{2})}ds_{1}\int_{s_{1}}^{\infty}\psi_{1}(s_{1},s_{2})e^{-s_{2}\Phi(\lambda_{2})}ds_{2}\nonumber\\
    &+&L_{2}(\lambda_{2},\lambda_{1})\int_{0}^{\infty}e^{-s_{1}\Phi(\lambda_{1})}ds_{1}\int_{0}^{s_{1}}\psi_{2}(s_{1},s_{2})e^{-s_{2}\Phi(\lambda_{1}+\lambda_{2})+s_{2}\Phi(\lambda_{1})}ds_{2}
    \label{xlxl}
\end{eqnarray}
where
$$
\psi_{1}(s_{1},s_{2})=\frac{2Ds_{1}}{\pi}\left[\pi-2\arctan\left(\sqrt{\frac{s_{2}}{s_{1}}-1}\right)+2\sqrt{\frac{s_{2}}{s_{1}}-1}\right],
$$
$$
\psi_{2}(s_{1},s_{2})=\frac{2Ds_{2}}{\pi}\left[\pi-2\arctan\left(\sqrt{\frac{s_{2}}{s_{2}}-1}\right)+2\sqrt{\frac{s_{1}}{s_{2}}-1}\right].
$$
Now we compute the integrals in the above expression. The first integral is trivial:
$$
\int_{0}^{\infty}e^{-s_{1}\Phi(\lambda_{2}+\lambda_{1})}s_{1}ds_{1}=\frac{1}{\Phi(\lambda_{2}+\lambda_{1})^{2}}.
$$
Then inner integral of the second term can be found integrating $\psi_1(s_1,s_2)$ term by term. To this end we need the following integral
$$
\int_{0}^{\infty}e^{-ax}\arctan(\sqrt{x})dx=2\int_{0}^{\infty}ye^{-ay^{2}}\arctan(y)dy=\frac{1}{a}\int_{0}^{\infty}\frac{e^{-ay^{2}}}{1+y^{2}}dy=\frac{\pi e^{a}}{2a}\textrm{erfc}(\sqrt{a})
$$
where we have integrated by parts in the second equality and made use of Eq. 3.466.1 in Ref. \cite{Gr15} in the third equality. Finally we obtain
\begin{eqnarray}
    \int_{s_{1}}^{\infty}\psi_{1}(s_{1},s_{2})e^{-s_{2}\Phi(\lambda_{2})}ds_{2}&=&\frac{2Ds_{1}}{\Phi(\lambda_{2})}e^{-s_{1}\Phi(\lambda_{2})}-\frac{2Ds_{1}}{\Phi(\lambda_{2})}\textrm{erfc}\left(\sqrt{s_{1}\Phi(\lambda_{2})}\right)\nonumber\\
    &+&\frac{2D}{\Phi(\lambda_{2})^{3/2}}\sqrt{\frac{s_{1}}{\pi}}e^{-s_{1}\Phi(\lambda_{2})}.
\end{eqnarray}
Computing the integral over $s_1$ of the above expression, the second term of  \eqref{xlxl} is
\begin{eqnarray*}
   & & L_{1}(\lambda_{2},\lambda_{1})\int_{0}^{\infty}e^{-s_{1}\Phi(\lambda_{2}+\lambda_{1})+s_{1}\Phi(\lambda_{2})}ds_{1}\int_{s_{1}}^{\infty}\psi_{1}(s_{1},s_{2})e^{-s_{2}\Phi(\lambda_{2})}ds_{2}\nonumber\\
   &=&\frac{2D\Phi(\lambda_{2})}{\lambda_{1}\lambda_{2}\left[\Phi(\lambda_{2}+\lambda_{1})-\Phi(\lambda_{2})\right]}\left\{ -\frac{2}{\Phi(\lambda_{2}+\lambda_{1})}+\frac{\Phi(\lambda_{2})}{\Phi(\lambda_{2}+\lambda_{1})^{2}}+\frac{\sqrt{\Phi(\lambda_{2}+\lambda_{1})}}{2\Phi(\lambda_{2})^{3/2}}\right.\\
   &+&\left.\frac{1}{2\sqrt{\Phi(\lambda_{2})\Phi(\lambda_{2}+\lambda_{1})}}\right\} 
\end{eqnarray*}
The third term of  \eqref{xlxl} is exactly equal to the above term but doing the permutation of indexes $1\leftrightarrow2$. Gathering the above results and introducing them into  \eqref{xlxl} we can express \eqref{alal} as
\begin{eqnarray}
  & &  \left\langle \tilde{\mathcal{A}}(\lambda_{1})\tilde{\mathcal{A}}(\lambda_{2})\right\rangle =\frac{2D}{\tau^{\alpha-1}\lambda_{1}\lambda_{2}}\left\{ \frac{1}{\lambda_{1}\lambda_{2}\left(\lambda_{1}+\lambda_{2}\right)^{\alpha}}+\frac{1}{2\lambda_{1}\lambda_{2}^{1+\alpha/2}\left(\lambda_{1}+\lambda_{2}\right)^{\alpha/2}}+\frac{1}{2\lambda_{2}\lambda_{1}^{1+\alpha/2}\left(\lambda_{1}+\lambda_{2}\right)^{\alpha/2}}\right.\nonumber\\
    &-&\left.\frac{1}{\lambda_{1}\lambda_{2}\left(\lambda_{1}+\lambda_{2}\right)^{\alpha/2}}\frac{1}{\lambda_{2}^{\alpha/2}+\left(\lambda_{1}
    +\lambda_{2}\right)^{\alpha/2}}-\frac{1}{\lambda_{1}\lambda_{2}\left(\lambda_{1}+\lambda_{2}\right)^{\alpha/2}}\frac{1}{\lambda_{1}^{\alpha/2}+\left(\lambda_{1}+\lambda_{2}\right)^{\alpha/2}}\right\}.
    \label{AlAl}
\end{eqnarray}
Now we have to compute the double Laplace inversion to \eqref{AlAl}. To do this we need to use the following Laplace inversion properties:
$$
\mathcal{L}_{\lambda\to t}^{-1}\left[\tilde{f}(\lambda+a)\right]=e^{-at}\mathcal{L}_{\lambda\to t}^{-1}\left[\tilde{f}(\lambda)\right]=e^{-at}f(t),\quad\mathcal{L}_{\lambda\to t}^{-1}\left[\frac{1}{\lambda^{\beta}}\right]=\frac{t^{\beta-1}}{\Gamma(\beta)}
$$
$$
\mathcal{L}_{\lambda\to t}^{-1}\left[\tilde{f}(\lambda)e^{-a\lambda}\right]=\theta(t-a)\mathcal{L}_{\lambda\to t-a}^{-1}\left[\tilde{f}(\lambda)\right],\quad\mathcal{L}_{\lambda\to t}^{-1}\left[\frac{\tilde{f}(\lambda)}{\lambda}\right]=\int_{0}^{t}f(t')dt'
$$
and the convolution property, we can show that
$$
\mathcal{L}_{\lambda_{1}\to t_{1}}^{-1}\mathcal{L}_{\lambda_{2}\to t_{2}}^{-1}\left[\frac{1}{\lambda_{1}^{\gamma}\lambda_{2}^{\beta}\left(\lambda_{1}+\lambda_{2}\right)^{\delta}}\right]=\frac{1}{\Gamma(\alpha)\Gamma(\beta)\Gamma(\gamma)}\int_{0}^{\min(t_{1},t_{2})}\tau_{1}^{\delta-1}(t_{1}-\tau_{1})^{\gamma-1}(t_{2}-\tau_{2})^{\beta-1}d\tau_{1}
$$
The inverse Laplace transform of the three first terms of the right hand side follows easily using the above formula. The last two terms are equal under the permutation $1\leftrightarrow2$. The Laplace inversion of one of these terms is
\begin{eqnarray*}
    & &\mathcal{L}_{\lambda_{1}\to t_{1}}^{-1}\mathcal{L}_{\lambda_{2}\to t_{2}}^{-1}\left[\frac{1}{\lambda_{1}^{2}\lambda_{2}^{2}\left(\lambda_{1}+\lambda_{2}\right)^{\alpha/2}}\frac{1}{\lambda_{1}^{\alpha/2}+\left(\lambda_{1}+\lambda_{2}\right)^{\alpha/2}}\right]\\
    &=&\mathcal{L}_{\lambda_{1}\to t_{1}}^{-1}\left\{ \frac{1}{\lambda_{1}^{2}}\int_{0}^{t_{2}}d\tau_{2}(t_{2}-\tau_{2})e^{-\lambda_{1}\tau_{2}}\mathcal{L}_{\lambda_{2}\to\tau_{2}}^{-1}\left[\frac{1}{\lambda_{1}^{\alpha/2}+\left(\lambda_{1}+\lambda_{2}\right)^{\alpha/2}}\right]\right\} 
\end{eqnarray*}
To compute the inner inverse Laplace transform we make use of the Laplace transform of the two-parametric Mittag-Leffler function (see Eq. 4.9.3 in \cite{Go20})
$$
\mathcal{L}_{t\to\lambda}\left[t^{\beta-1}E_{\sigma,\beta}\left(at^{\sigma}\right)\right]=\frac{\lambda^{\sigma-\beta}}{\lambda^{\sigma}-a}
$$
so that
\begin{eqnarray*}
   & & \mathcal{L}_{\lambda_{1}\to t_{1}}^{-1}\mathcal{L}_{\lambda_{2}\to t_{2}}^{-1}\left[\frac{1}{\lambda_{1}^{2}\lambda_{2}^{2}\left(\lambda_{1}+\lambda_{2}\right)^{\alpha/2}}\frac{1}{\lambda_{1}^{\alpha/2}+\left(\lambda_{1}+\lambda_{2}\right)^{\alpha/2}}\right]\\
   &=&\int_{0}^{t_{2}}d\tau_{2}(t_{2}-\tau_{2})\tau_{2}^{\alpha-1}\theta(t_1-\tau_2)\mathcal{L}_{\lambda_{1}\to t_{1}-\tau_2}^{-1}\left[\frac{1}{\lambda_{1}^{2}}E_{\frac{\alpha}{2},\alpha}\left(-\lambda_{1}^{\alpha/2}\tau_{2}^{\alpha/2}\right)\right].
\end{eqnarray*}
To perform the inverse Laplace transform with respect to $\lambda_1$ we make use of Eq. 1.50 in \cite{Mat09} which relates the Mittag-Leffler function to the Fox function. Using now Eq. 2.21 in \cite{Mat09} to compute the inverse Laplace transform of the Fox function we find
\begin{eqnarray}
    & &\mathcal{L}_{\lambda_{1}\to t_{1}}^{-1}\mathcal{L}_{\lambda_{2}\to t_{2}}^{-1}\left[\frac{1}{\lambda_{1}^{2}\lambda_{2}^{2}\left(\lambda_{1}+\lambda_{2}\right)^{\alpha/2}}\frac{1}{\lambda_{1}^{\alpha/2}+\left(\lambda_{1}+\lambda_{2}\right)^{\alpha/2}}\right]\nonumber\\
    &=&\int_{0}^{\min(t_{1},t_{2})}d\tau_{2}(t_{2}-\tau_{2})(t_{1}-\tau_{2})\tau_{2}^{\alpha-1}H_{2,2}^{1,1}\left[\left(\frac{\tau_{2}}{t_{1}-\tau_{2}}\right)^{\frac{\alpha}{2}}\left|\begin{array}{cc}
(0,1), & \left(2,\frac{\alpha}{2}\right)\\
(0,1), & \left(1-\alpha,\frac{\alpha}{2}\right)
\end{array}\right.\right].
\label{l1l1}
\end{eqnarray}
Introducing the above results into \eqref{AlAl} and using Eq. 2.51 in \cite{Mat09} to compute the integrals of the Fox functions we finally find, after doing $t_1=t_2=t$
\begin{eqnarray}
   \left\langle \mathcal{A}(t)^{2}\right\rangle =\frac{2D}{\tau^{\alpha-1}}t^{\alpha+2}\left\{ \frac{8+\alpha}{2\Gamma(3+\alpha)}-2\int_{0}^{\infty}\frac{d\chi}{\chi^{1-\alpha}(1+\chi)^{3+\alpha}}H_{2,2}^{1,1}\left[\chi^{\frac{\alpha}{2}}\left|\begin{array}{cc}
(0,1), & \left(2,\frac{\alpha}{2}\right)\\
(0,1), & \left(1-\alpha,\frac{\alpha}{2}\right)
\end{array}\right.\right]\right\} .
\label{A2sf}
\end{eqnarray}
To compute the integral of the Fox function we note that it can be regarded as the Mellin transform of the product of two Fox functions if convert the factor $(1+\chi)^{-3-\alpha}$ into a Fox function. Using Eq. 2.9.6 in Ref \cite{Ki04} we can write
$$
\frac{1}{(1+\chi)^{3+\alpha}}=\frac{1}{\Gamma(3+\alpha)}H_{1,1}^{1,1}\left[\chi\left|\begin{array}{c}
(-2-\alpha,1)\\
(0,1)
\end{array}\right.\right]
$$
and from Eq. 2.8.12 in Ref \cite{Ki04} we find
\begin{eqnarray}
  & &  \int_{0}^{\infty}\frac{d\chi}{\chi^{1-\alpha}(1+\chi)^{3+\alpha}}H_{2,2}^{1,1}\left[\chi^{\frac{\alpha}{2}}\left|\begin{array}{cc}
(0,1), & \left(2,\frac{\alpha}{2}\right)\\
(0,1), & \left(1-\alpha,\frac{\alpha}{2}\right)
\end{array}\right.\right]\nonumber\\
&=&\frac{1}{\Gamma(3+\alpha)}\int_{0}^{\infty}d\chi\chi^{\alpha-1}H_{2,2}^{1,1}\left[\chi^{\frac{\alpha}{2}}\left|\begin{array}{cc}
(0,1), & \left(2,\frac{\alpha}{2}\right)\\
(0,1), & \left(1-\alpha,\frac{\alpha}{2}\right)
\end{array}\right.\right]H_{1,1}^{1,1}\left[\chi\left|\begin{array}{c}
(-2-\alpha,1)\\
(0,1)
\end{array}\right.\right]\nonumber\\
&=&\frac{1}{\Gamma(3+\alpha)}H_{3,3}^{2,2}\left[1\left|\begin{array}{ccc}
(0,1), & \left(1-\alpha,\frac{\alpha}{2}\right), & \left(2,\frac{\alpha}{2}\right)\\
(0,1), & \left(3,\frac{\alpha}{2}\right), & \left(1-\alpha,\frac{\alpha}{2}\right)
\end{array}\right.\right].
\label{H1}
\end{eqnarray}
To compute \eqref{H1} we convert the Fox function into a power series using Eq. 1.3.7 in Ref. \cite{Ki04}. For a general argument $0<\chi<1$ 
$$
H_{3,3}^{2,2}\left[\chi\left|\begin{array}{ccc}
(0,1), & \left(1-\alpha,\frac{\alpha}{2}\right), & \left(2,\frac{\alpha}{2}\right)\\
(0,1), & \left(3,\frac{\alpha}{2}\right), & \left(1-\alpha,\frac{\alpha}{2}\right)
\end{array}\right.\right]=\sum_{k=0}^{\infty}(-\chi)^{k}\left(2-k\frac{\alpha}{2}\right)=\frac{1}{1+\chi}\left(2+\frac{\alpha}{2}\frac{\chi}{1+\chi}\right)
$$
so that the evaluation at $\chi=1$ is an analytical continuation and the sum must be interpreted as an Abel sum. Then,
$$
H_{3,3}^{2,2}\left[1\left|\begin{array}{ccc}
(0,1), & \left(1-\alpha,\frac{\alpha}{2}\right), & \left(2,\frac{\alpha}{2}\right)\\
(0,1), & \left(3,\frac{\alpha}{2}\right), & \left(1-\alpha,\frac{\alpha}{2}\right)
\end{array}\right.\right]=1+\frac{\alpha}{8}
$$
and from \eqref{H1} Eq. \eqref{A2sf} turns into
$$
\left\langle \mathcal{A}(t)^{2}\right\rangle =\frac{4K_{\alpha}}{\Gamma\left(3+\alpha\right)}\left(1+\frac{\alpha}{8}\right)t^{2+\alpha}.
$$

\section{Derivation of Eqs. \eqref{aam1} and \eqref{aam2}} \label{app:aam1_aam2}
The first moment of the absolute area can be found from  \eqref{aa1m} using the change of variables \eqref{y(x)}. So that,
$$
\left\langle \mathcal{A}(t)^{2}\right\rangle =2\int_{0}^{t}d\tau\int_{0}^{\infty}xP(x,\tau)dx=2\left((2-\alpha)\sqrt{\frac{D_{0}}{2}}\right)^{\frac{2}{2-\alpha}}\int_{0}^{t}d\tau\int_{0}^{\infty}y^{\frac{2}{2-\alpha}}P(y,\tau)dy
$$
where $P(y,\tau)$ is given in \eqref{oty}. Performing the integrals one readily finds \eqref{aam1}.
Analogously, for the second moment we make use of \eqref{aeqcA} to write
$$
\left\langle \mathcal{A}(t)^{2}\right\rangle =2\left(\frac{2-\alpha}{\sqrt{2}}\sqrt{D_{0}}\right)^{\frac{4}{2-\alpha}}\int_{0}^{t}d\tau_{2}\int_{0}^{\tau_{2}}d\tau_{1}\int_{-\infty}^{\infty}|y_{2}|{}^{\frac{2}{2-\alpha}}dy_{2}\int_{-\infty}^{\infty}|y_{1}|{}^{\frac{2}{2-\alpha}}dy_{1}P(y_{2},\tau_{2}|y_{1},\tau_{1})P(y_{1},\tau_{1}).
$$
The integral over $y_2$ can be performed following the sames steps as in \ref{app:A2dx}. We find
\begin{eqnarray*}
 & &\int_{-\infty}^{\infty}|y_{2}|^{\frac{2}{2-\alpha}}P(y_{2},\tau_{2}|y_{1},\tau_{1})dy_{2}\\
 &=&  \frac{\left(\tau_{2}-\tau_{1}\right)^{\frac{1}{2-\alpha}}}{\sqrt{2\pi}}\Gamma\left(\frac{4-\alpha}{2-\alpha}\right)e^{-\frac{y_{1}^{2}}{4(\tau_{2}-\tau_{1})}}\left[D_{-\frac{4-\alpha}{2-\alpha}}\left(-\frac{y_{1}}{\sqrt{\tau_{2}-\tau_{1}}}\right)+D_{-\frac{4-\alpha}{2-\alpha}}\left(\frac{y_{1}}{\sqrt{\tau_{2}-\tau_{1}}}\right)\right].
\end{eqnarray*}
Using 
$$
D_{-\nu}(-z)+D_{-\nu}(z)=\frac{\sqrt{\pi}2^{\frac{2-\nu}{2}}}{\Gamma((\nu+1)/2)}e^{-\frac{z^{2}}{4}}M\left(\frac{\nu}{2},\frac{1}{2},\frac{z^{2}}{2}\right)
$$
and again Eq. 7.621.4 of Ref. \cite{Gr15} we find
\begin{eqnarray*}
  &  &\int_{-\infty}^{\infty}|y_{2}|{}^{\frac{2}{2-\alpha}}dy_{2}\int_{-\infty}^{\infty}|y_{1}|{}^{\frac{2}{2-\alpha}}dy_{1}P(y_{2},\tau_{2}|y_{1},\tau_{1})P(y_{1},\tau_{1})\\
  &=&\frac{1}{\sqrt{\pi}}\frac{\Gamma\left(\frac{4-\alpha}{2-\alpha}\right)\Gamma\left(\frac{4-\alpha}{4-2\alpha}\right)}{\Gamma\left(\frac{3-\alpha}{2-\alpha}\right)}\left(1-\frac{\tau_{1}}{\tau_{2}}\right)^{\frac{6-\alpha}{2(2-\alpha)}}\left(\tau_{1}\tau_{2}\right)^{\frac{1}{2-\alpha}}{}_{2}F_{1}\left(\frac{3-\alpha}{2-\alpha},\frac{3-\alpha}{2-\alpha};\frac{3}{2},\frac{\tau_{1}}{\tau_{2}}\right)
\end{eqnarray*}
Now we have to integrate the above expression over $\tau_1$ and $\tau_2$ to compute the second moment. Using again Eq. 7.512.5 of Ref. \cite{Gr15} we finally find \eqref{aam2}.

\section{Numerical Simulations Methods}\label{app:simulations}

In this Appendix, we describe the numerical algorithms employed to obtain the simulation results presented in the main text. All numerical data are generated from Monte Carlo simulations of ensembles comprising $N$ independent particle trajectories.

SBM and HBM are simulated using the Euler-Maruyama discretization scheme.  Within this, a single step of the stochastic trajectory is characterized by a time increment $\tau_i$ and a spatial increment $\Delta_i$. We fix $\tau_i = dt$ and define
\begin{equation}
\Delta_i = \sqrt{2D\,dt}\,\xi,
\end{equation}
where $\xi$ is a Gaussian random variable drawn from a normal distribution $\mathcal{N}(0,1)$. For SBM, the diffusion coefficient is explicitly time-dependent and given by Eq. \eqref{Dsbm}. For HBM, the diffusion coefficient depends on the particle position but to prevent numerical instabilities in the vicinity of $x = 0$, we regularize the diffusion coefficient \cite{ChMe13} as $D(x) = D_0\left(|x|^{\alpha} + \varepsilon\right)$ with $\varepsilon = 10^{-4}$.
FBM trajectories are generated using the circulant embedding method (a detailed description of this algorithm can be found in Ref. \cite{perrin2002fast}). Subdiffusive-CTRW trajectories are generated from a sequence of independent waiting times $t_i$ and spatial jumps $x_i$. During each waiting time $t_i$, the particle remains at a fixed position, after which it performs a spatial jump. The jump lengths are chosen as $x_i = \pm l$ (with equal probability). The waiting times are sampled from a Mittag-Leffler PDF, as defined in Eq. \eqref{ml} of the main text. The diffusion  coefficient is $D=l^2/2 \tau$. 

Throughout the main text numerical simulations, the model parameters were fixed as follows. For the SBM, we used $dt = 0.1$, $K = 1$, and $N = 2 \times 10^{4}$. 
For the fBM, the parameters were $dt = 1$, $D_H = 1$ and $N = 10^{4}$. 
For the subdiffusive-CTRW, the jump length was set to $l = 1$,  $\tau = 100$ (so that $D=1/200$), and $N = 10^{5}$.  Finally, for the HBM, we fixed $dt = 0.1$,  $D_0 = 1$, $\epsilon = 10^{-3}$, and $N = 3 \times 10^{4}$.

\section{PDFs for fBM and HBM}\label{app:figures}

\begin{figure}[h]
    \centering
    \includegraphics[width=0.99\linewidth]{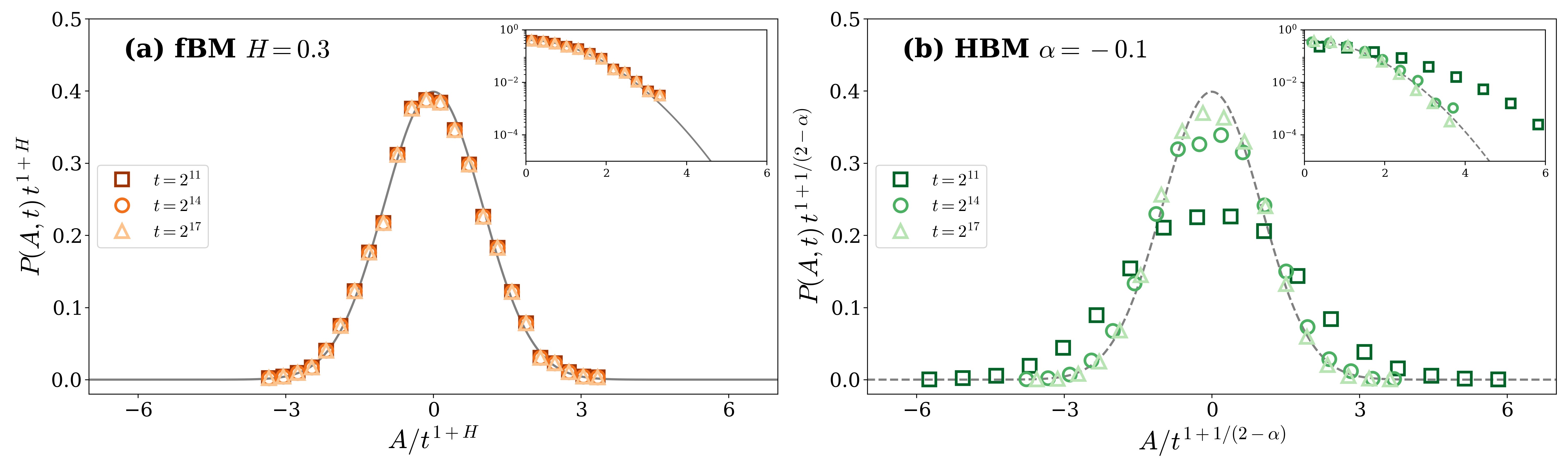}
    \caption{ Probability distribution of $P(A,t)$ vs $A$ at different times $t$ . (a) Fractional Brownian Motion (fBM). The dotted line corresponds to the prediction given if the functional was Gaussian. (b) Heterogeneous Brownian motion (HBM). The dotted line corresponds to the prediction given if the functional was Gaussian. The simulations details are provided in \ref{app:simulations}.}
    \label{fig:pdf_A_app}
\end{figure}

\begin{figure}[h]
    \centering
    \includegraphics[width=0.99\linewidth]{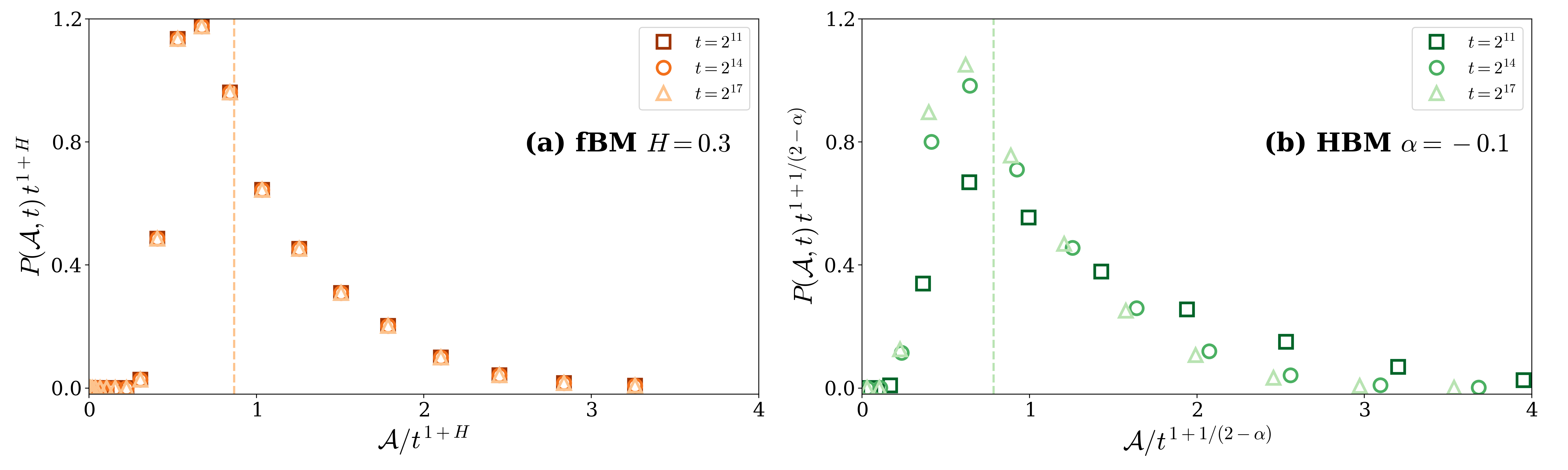}
    \caption{ Probability distribution of $P(\mathcal{A,t})$ vs $\mathcal{A}$ at different times $t$ . (a) Fractional Brownian Motion (fBM). The vertical line corresponds to the mean value predicted by Eq. \eqref{a1a2aa}. (b) Heterogeneous Brownian motion (HBM). The vertical line corresponds to the mean value predicted by Eq. \eqref{aam1}. The simulations details are provided in \ref{app:simulations}. }
   \label{fig:pdf_Aabs_app}
\end{figure}

For completeness, we present here the PDFs of the area $A$ (Fig. \ref{fig:pdf_A_app}) and of the absolute area $\mathcal{A}$ (Fig. \ref{fig:pdf_Aabs_app}) for both the fBM and HBM. Again, we find that the scaling Eq. \eqref{eq:scaling} is fulfilled. Moreover, since the fBM is a Gaussian process, we also see that $P(A,t)$ is a Gaussian, and follows Eq. \eqref{pdf_fbm}, as expected. Finally, for the heterogeneous case, we find that the convergence to the results is slower, as already seen for the moments and EB.

\bibliographystyle{elsarticle-num}
\bibliography{main}

\end{document}